\documentclass[journal]{IEEEtran}

\usepackage{epsfig,color,amsmath,cite}
\usepackage{amsthm} 
\usepackage{amsmath}    
\IEEEoverridecommandlockouts
\usepackage{empheq}
\usepackage{bm}
\usepackage{epstopdf}
\usepackage{amssymb}
\usepackage{url}
\usepackage{enumitem} 
\usepackage{multirow}
\usepackage{hhline}
\usepackage{booktabs}
\usepackage{mathtools}
\usepackage{makecell}
\usepackage[linesnumbered,boxed,commentsnumbered,ruled,vlined,longend]{algorithm2e}
\usepackage{comment}

\DeclareMathOperator*{\minimize}{minimize}

\DeclareMathOperator*{\subjectto}{subject\ to}

\makeatother
\DeclareMathAlphabet\mathbfcal{OMS}{cmsy}{b}{n}



\usepackage{stackengine}

\newcommand{\mat}[1]{\boldsymbol{#1}}

\newcommand{\bmat}[1]{\begin{bmatrix} #1 \end{bmatrix}}

\providecommand{\mA}{\ensuremath{\mat{A}}}
\providecommand{\mB}{\ensuremath{\mat{B}}}
\providecommand{\mC}{\ensuremath{\mat{C}}}
\providecommand{\mD}{\ensuremath{\mat{D}}}
\providecommand{\mE}{\ensuremath{\mat{E}}}
\providecommand{\mF}{\ensuremath{\mat{F}}}

\providecommand{\mH}{\ensuremath{\mat{H}}}
\providecommand{\mI}{\ensuremath{\mat{I}}}

\providecommand{\mK}{\ensuremath{\mat{K}}}

\providecommand{\mO}{\ensuremath{\mat{O}}}
\providecommand{\mP}{\ensuremath{\mat{P}}}
\providecommand{\mQ}{\ensuremath{\mat{Q}}}

\providecommand{\mS}{\ensuremath{\mat{S}}}
\providecommand{\mT}{\ensuremath{\mat{T}}}
\providecommand{\mU}{\ensuremath{\mat{U}}}
\providecommand{\mV}{\ensuremath{\mat{V}}}

\providecommand{\mX}{\ensuremath{\mat{X}}}
\providecommand{\mY}{\ensuremath{\mat{Y}}}

\newcommand{\m}{\boldsymbol}
\allowdisplaybreaks[4]
\usepackage[colorlinks = true,
linkcolor = blue,
urlcolor  = blue,
citecolor = blue,
anchorcolor = blue]{hyperref}

\newcommand{\mc}[1]{\mathcal{#1}}
\newcommand{\mbb}[1]{\mathbb{#1}}
\newcommand{\mr}[1]{\mathrm{#1}}
\usepackage[framemethod=TikZ]{mdframed}
\mdfdefinestyle{MyFrame}{%
	linecolor=black,
	outerlinewidth=1.25pt,
	roundcorner=1.25pt,
	innerrightmargin=5pt,
	innerleftmargin=5pt,}
	
\usepackage[noabbrev]{cleveref}

\usepackage{mathtools}

\DeclarePairedDelimiter\abs{\lvert}{\rvert}%
\DeclarePairedDelimiter\norm{\lVert}{\rVert}%

\makeatletter
\let\oldabs\abs
\def\abs{\@ifstar{\oldabs}{\oldabs*}}
\let\oldnorm\norm
\def\norm{\@ifstar{\oldnorm}{\oldnorm*}}
\makeatother

\usepackage[english]{babel}
\usepackage[utf8]{inputenc}
\usepackage[super]{nth}

\usepackage{graphicx}
\usepackage{float}
\usepackage[caption = false]{subfig}

\usepackage{array}
\usepackage{threeparttable}

\usepackage[english]{babel}
\usepackage[utf8]{inputenc}
\usepackage[super]{nth}

\RequirePackage{filecontents}

\SetKwRepeat{Do}{do}{while}%

\usepackage{mleftright}

\usepackage{lipsum}

\title{\vspace{0.5cm}\centering \Large \textsc{{\textbf{Sorta Solving the OPF by \textit{Not} Solving the OPF:\\ DAE Control Theory and the Price of Realtime Regulation}}}}
\vspace{0.4cm}\author{Muhammad Nadeem and Ahmad F. Taha \vspace{-0.00008cm}
	\thanks{
	The authors are with Department of  Civil and Environmental Engineering, Vanderbilt University, Nashville, TN 37235. Taha has a secondary appointment  with the Electrical and Computer Engineering Department.
		Emails: muhammad.nadeem@vanderbilt.edu, ahmad.taha@vanderbilt.edu. This work is supported by the National Science Foundation under Grant ECCS 2151571 and CMMI 2152450. The authors would like to acknowledge Sebastian Nugroho for his help with the codes.}
}

\begin{document}

\newdimen\origiwspc%
\newdimen\origiwstr%
\origiwspc=\fontdimen2\font
\origiwstr=\fontdimen3\font

\fontdimen2\font=0.62ex

\maketitle

\begin{abstract}
This paper presents a new approach to approximate the AC optimal power flow (ACOPF). By eliminating the need to solve the ACOPF every few minutes, the paper showcases how a realtime feedback controller can be utilized in lieu of ACOPF and its variants. By \textit{(i)} forming the grid dynamics as a system of differential-algebraic equations (DAE) that naturally encode the non-convex OPF power flow constraints, \textit{(ii)} utilizing DAE-Lyapunov theory, and \textit{(iii)} designing a feedback controller that captures realtime uncertainty while being uncertainty-unaware, the presented approach demonstrates promises of obtaining solutions that are close to the OPF ones without needing to solve the OPF. The proposed controller responds in realtime to deviations in renewables generation and loads, guaranteeing improvements in system transient stability, while always yielding approximate solutions of the ACOPF with no constraint violations. As the studied approach herein yields slightly more expensive realtime generator controls, the corresponding price of realtime control and regulation is examined. Cost comparisons with the traditional ACOPF are also showcased---all via case studies on standard power networks.
\end{abstract}


\vspace{-0.05cm}

\vspace{-0.05cm}
\section{Introduction and Paper Contributions}
 \IEEEPARstart{I}{t}  is not an overstatement that the OPF problem---and its many variants---is arguably the most researched and solved optimization problem in the world. OPF~\cite{dommel1968optimal} refers to computing setpoints of generators in a power network every few minutes, allowing generation to meet the varying demand. In short, the problem minimizes the cost of generation from mostly fossil fuel-based power plants subject to power balance in transmission power lines (acting as equality constraints or $\m h(\m x) = \m 0$) and thermal line, voltages, and generation limits (acting as inequality constraints or $\m g(\m x) \leq \m 0$). This optimization problem can be written as 
\begin{equation}\label{equ:OPF-HL}
	\hspace{-0.25cm}	\textbf{OPF:}\; \minimize \; f(\m x) \; \subjectto \; \m g(\m x) \hspace{-0.05cm}\leq\hspace{-0.05cm} \m 0, \; \m h(\m x)\hspace{-0.05cm}	 =\hspace{-0.05cm} \m 0.
\end{equation}
Due to the nonconvexity in the power balance equality constraints $\m h(\m x) = \m 0$, the OPF is infamously non-convex. The infamy is \textit{not} because the nonconvexity is too insufferable however insufferability is defined; it is because OPF has become a textbook example of practical optimization problems in operations research and systems engineering.

\noindent \textbf{Brief Literature Review. } In pursuit of overcoming this nonconvexity, hundreds of papers yearly investigate methods to solve variants of OPF. The OPF can also make you a millionaire: The US department of energy and ARPA-E have a competition, called grid optimization (GO) Competition~\cite{aravena2022recent,holzer2021grid}, where academics and practitioners compete in solving variants of the OPF with up to \$3 million in prizes. To solve the OPF, academics often resort to one of these four approaches.  \textit{(i)} Assume DC power flow and eliminate some variables, resulting in convex quadratic programs that can be solved efficiently for large power systems \cite{taylor2015convex,MomohITPWRS1999, FDR, ArdakaniITPWRS2013}. \textit{(ii)} Derive semidefinite programming (SDP) relaxations of OPF appended with methods to recover an optimal solution \cite{AndersenITPWRS2014,Louca, LavaeiITPWRS2012,MolzahnITPWRS, MolzahnITPWRS2013,LowITCNS2014,MadanITPWRS2015,MadaniITPWRS2016}. \textit{(iii)} Design global optimization methods with some performance guarantees under various relaxations of nonconvex OPF \cite{gopalakrishnan2012global,lu2018tight, LeeITPWRS2020}.  \textit{(iv)} Obtain machine learning-based algorithms that learn solutions to OPF \cite{baker2019learning,HuangITPWRS2022,PanITPWRS2021,Zamzam2020, baker2022emulating,ZhouJMPSCE2020}.  A thorough description of the OPF literature is outside the scope of this work.

Relevant to these approaches that \textit{only} focus on ACOPF are methods and algorithms that study stability-constrained OPF where dynamic stability or optimal control metrics are appended to the OPF problem \cite{BazrafshanITSG2019, LiITCNS2016,DorflerITCNS216, TRIP2016240}. This integration of system \textit{operating cost} and \textit{dynamic stability} results in the merging of power system's secondary and tertiary control layers. This means that the five-minute generator setpoints provided by the tertiary layer not only minimize the system operating cost but also allow the system to be more controllable or more stable. However, this does not circumvent the issues with the nonconvex equality constraints in the ACOPF problem and still convex relaxation and linearization are required. This direction of stability or control-constrained OPF, while seemingly distinct, aligns closely with the method we propose in our approach.


Furthermore, due to the increasing penetration of renewable energy resources, complex load demands, and other power electronics-based devices in the future power grid, the 5--10 minutes setpoints provided by the traditional OPF may not be valid/optimal because of the time-separation and slow update process \cite{TangITSG2017}. With that in mind, this paper investigates a new approach of solving the OPF problem in realtime. This is done in a way by virtually ignoring~\eqref{equ:OPF-HL} and dumping the OPF problem into a feedback control problem that inherently satisfies the constraint set in~\eqref{equ:OPF-HL}, while simultaneously performing other tasks such as load frequency regulation and realtime control. Next, we explain in detail how this approach works.  A more detailed discussion of the literature is omitted from here (for brevity and clarity) and instead is given after the formulation of the proposed controller (Section~\ref{sec:NDAE_control}). 


\noindent \textbf{Main Idea. }First, formulate a dynamic, differential algebraic equation (DAE) model of power systems. This model incorporates algebraic equations that model power flows (the nonconvex constraints in~\eqref{equ:OPF-HL}) as well as generator dynamics. Then, solve a control problem that computes \textit{(a)} OPF setpoints (i.e., generator output power) and \textit{(b)} their deviations in a feedback fashion by utilizing phasor measurement units  (PMUs) data in realtime, while forgoing the need to solve for the power flow variables---yet still ensuring that physical constraints are not violated. By formulating a realtime, feedback-driven control problem that solves for a time-invariant feedback gain matrix, and feeding that gain into a highly scalable differential algebraic equation (DAE) solver, we avoid actually having to solve a nonconvex optimization problem for the OPF variables. 

In short, we address a robust feedback control problem for the power system model, which not only incorporates the key constraints found in the OPF problem but also delivers a solution closely aligned with the OPF results. This approach ensures optimal generator setpoints akin to those in the OPF, while also enhancing system transient stability by introducing damping to system oscillations.

The presented approach in this paper does \textit{not} actually solve the OPF with the generator's cost curves---but as observed later in the paper, we showcase using thorough simulations how the proposed approach is very close to the optimality under transient conditions similar to the OPF solution. Furthermore, in a real-world setting, the DAE solver is replaced with the actual system meaning that even the DAEs do not have to be simulated. The DAEs already encode the nonconvex constraints (using constant system matrices) in OPF---a key factor in the proposed method.  

\noindent \textbf{Paper Contributions.} The presented approach in this paper is endowed with the following key properties and contributions. The OPF-controller: \textit{(i)} Circumvents the need to solve or deal with the nonconvex equality constraints modeling power flow. That is, we find generators' setpoints in realtime while knowing that these setpoints do satisfy and abide by the power flow constraints.
\textit{(ii)} Deals with the uncertainty in renewables, loads, and parameters in a control-theoretic way. In contrast with vintage robust optimization or the more intricate distributionally robust optimization algorithms, the developed approach here is truly uncertainty unaware. \textit{(iii)} Utilizes realtime information from grid sensors such as PMU via state estimators allowing for realtime micro-adjustments of dispatchable generation. This is in contrast with OPF formulations that do not utilize grid measurements for better dispatch of generators. \textit{(iv)} Eliminates the need to separate the OPF and the secondary control time scales: this approach serves the purposes of both OPF and control, so the need to separate the two becomes unnecessary.  \textit{(v)} Seamlessly incorporates advanced models of renewables, resulting in setpoints for fuel-based generators that are aware of the dynamics of solar and wind farms, etc. 

\noindent {\bf Notations.} Bold lowercase and uppercase letters represent vectors and matrices respectively, while all calligraphic letters denote sets such as $\mathcal{R}$, $\mathcal{N}$, etc. The set of real-valued $x$ by $y$ matrices is represented as $\mathbb{R}^{x \times y}$. Similarly positive definite matrix of size $x$ by $y$ is denoted as $\mathbb{S}_{++}^{x \times y}$. We represent identity and zero matrices of appropriate dimension as $\mI$ and $\mO$. For any matrix $\mA$, symbols $\mA^{\perp}$, $\mA^T$, $\|\mA\|_0$, $\bar{\sigma}[\mA]$, and $\|\mA\|_2$  denote its, orthogonal complement, transpose, total number of non-zero elements, largest singular value, and $\mathcal{L}_2$-norm, respectively. We represent positive/negative definiteness as $\succ 0$/$\prec 0$ and positive semi-definiteness by $\succeq 0$. Symbol $*$ represents the symmetric elements in a given symmetric matrix. We denote the union (combination) of two sets via symbol $\cup$ such as $\mathcal{G} \cup \mathcal{R}$.  We use $\mr{diag}$ to show a diagonal matrix. The set $\mathbb{R}^x$ represents a column vector of $x$ elements and $\mathbb{R}_{++}$ represents a positive scalar. Given a vector $a(t)$ in time interval $t \in [0,\infty)$, its $\mathcal{L}_2$-norm is represented as $\sqrt{\int_0^\infty \|a(t)\|^2 dt}$. Also, for the sake of simplicity, we omit time dependency, i.e., $(t)$ in representing some of the time-dependent vectors. 

\noindent \textbf{Paper Organization.} The remainder of the paper is organized as follows: Section \ref{sec:ACOPF} summarizes the ACOPF problem formulation. Section \ref{sec:modeling} presents the multi-machine NDAE model of power networks. Section \ref{sec:NDAE_control} explains the proposed methodology and its mathematical derivation. Numerical case studies are performed in Section \ref{sec:casestudies} while the paper is concluded in Section \ref{sec:conclusion}.

\vspace{-0.3cm}

\section{ACOPF Formulation}\label{sec:ACOPF}
In this section, we briefly present the ACOPF formulation.\footnote{We use ACOPF and OPF interchangeably in this paper.} We consider a power network consisting $N$ number of buses, modeled by a graph $(\mathcal{N},\mathcal{E})$ where $\mathcal{N}$ is the set of nodes and $\mathcal{E}$ is the set of edges. 
Note that $\mathcal{N}$ consists of traditional synchronous generator, renewable energy resources, and load buses, i.e.,  $\mathcal{N} = \mathcal{G} \cup \mathcal{R} \cup \mathcal{L}$ where $\mathcal{G}$ collects $G$ generator buses, $\mathcal{R}$ collects the buses containing $R$ renewables, while $\mathcal{L}$ collects $L$ load buses. 
The generator's supplied (real and reactive) power is denoted by $(P_{\mr{G}i},Q_{\mr{G}i})$ for bus $i \in \mc{G}$, and the bus voltages are depicted as $v_i$. The bus angle is represented as $\theta_i$ and the angle difference in a line is  $\theta_{ij}:= \theta_i-\theta_j$. The parameters $(G_{ij},B_{ij})$ respectively denote the conductance and susceptance between bus $i$ and $j$ which can be directly obtained from the network's bus admittance matrix \cite{sauer2017power}. Furthermore, quantities $(P_{\mr{R}i},Q_{\mr{R}i})$ denote the active and reactive power generated by renewables for bus $i \in \mc{R}$, while $(P_{\mr{L}i},Q_{\mr{L}i})$ denote the active and reactive power consumed by the loads for bus $i \in \mc{L}$. Essentially, renewables are modeled as negative loads. If a bus does not have generation, load, or a renewable source attached to it, the corresponding active/reactive powers are equal to zero.

Given the above notation, the ACOPF can be written as~\cite{taylor2015convex}
\begin{subequations}~\label{equ:ACOPF}
	\begin{eqnarray}
		\min_{\m P_{\mr{G}}, \m Q_{\mr{G}}, \m \theta, \m v} &&\hspace{-0.2cm} J_{\mr{OPF}}(\m P_\mr{G})\hspace{-0.09cm}=\hspace{-0.09cm} \sum_{i\in \mc{G}}a_i P_{\mr{G}i}^2 + b_i P_{\mr{G}i}  + c_i  \\
		\subjectto && \forall i \in \mathcal{N}:  P_{\mr{G}i} + P_{\mr{R}i}	+P_{\mr{L}i} = \notag\\
		&&   v_i \sum_{j=1}^{N} v_j\hspace{-0.05cm}\left(G_{ij}\cos \theta_{ij} +B_{ij}\sin \theta_{ij}\right)  \label{equ:PF1} \\ 
		&&	\forall i \in \mathcal{N}:	Q_{\mr{G}i} + Q_{\mr{R}i}	+Q_{\mr{L}i}  = \notag \\
		&& v_i \sum_{j=1}^{N} v_j\left(G_{ij}\sin \theta_{ij} - B_{ij}\cos \theta_{ij}\right) \label{equ:PF2}\\
		&& \forall i \in \mathcal{G}: P_{\mr{G}i}^{\min} \leq P_{\mr{G}i} \leq P_{\mr{G}i}^{\max} \label{equ:PG_constratints}\\
		&& \forall i \in \mathcal{G}: Q_{\mr{G}i}^{\min} \leq Q_{\mr{G}i} \leq Q_{\mr{G}i}^{\max} \label{equ:QG_constraints}\\
		&&  \forall i \in \mathcal{N}: v_{i}^{\min} \leq v_i \leq v_{i}^{\max}
		\\
		&&  \forall i \in \mathcal{N}: S_{f_i} \leq F_\mr{max}
		\\
		&&  \forall i \in \mathcal{N}: S_{t_i} \leq F_\mr{max}.
	\end{eqnarray}
\end{subequations}
The variables in the ACOPF are the active/reactive powers for generator buses and angles and voltages for all buses $(\m P_{\mr{G}}, \m Q_{\mr{G}}, \m \theta, \m v)$. In~\eqref{equ:ACOPF}, the objective function $J_{\mr{OPF}}(\m P_\mr{G})$ minimizes the generator's convex quadratic cost function with parameters $a_i, b_i,$ and $c_i$. The first two constraints model power flow balance in the network---a nonlinear, non-convex relation between the variables. The last five constraints represent upper and lower bounds on the generators' power as well as bus voltages and line flow constraints, with $S_{f_i}$, $S_{t_i}$ representing \textit{from} and \textit{to} line flows and $F_\mr{max}$ denoting maximum rating of the transmission lines. 

The ACOPF is usually solved every $5-10$ minutes, although the frequency at which it is solved depends on the computational power and updated predictions of renewables and loads. Ideally, a system operator would have all of the constraints satisfied at each time step $t$, and one would solve a realtime ACOPF that satisfies all constraints while optimizing the cost function. 

In the next section, we present the dynamics of the same power system with a focus on the realtime control problem. We then showcase that the proposed realtime controller inherently satisfies some of the key ACOPF constraints. 
\section{Dynamics of Multi-Machine Power Systems }\label{sec:modeling}
Here, we describe the transient dynamics of a power system which by definition encodes the algebraic constraints~\eqref{equ:PF1} and~\eqref{equ:PF2}.  For the same power network, we can write the \nth{4}-order dynamics of synchronous generators as~\cite{sauer2017power}:
\begin{subequations} \label{eq:SynGen}
	\begin{align}
		\dot{\delta}_{i} &= \omega_{i} - \omega_{0} \label{eq:SynGen1} \\ 
		\begin{split}
			M_{i}\dot{\omega}_{i} &= T_{\mr{M}i}-P_{\mr{G}i}- D_{i}(\omega_{i}-\omega_{0}) \end{split}\label{eq:SynGen2}    \\ 
		T'_{\mr{d0}i}\dot{E}'_{i} &= -\tfrac{x_{\mr{d}i}}{x'_{\mr{d}i}}E'_{i} +\tfrac{x_{\mr{d}i}-x'_{\mr{d}i}}{x'_{\mr{d}i}}v_i\cos(\delta_{i}-\theta_i) + E_{\mr{fd}i}  \label{eq:SynGen3} \\
		T_{\mr{CH}i}\dot{T}_{Mi} &= -T_{\mr{M}i} - \tfrac{1}{R_{\mr{D}i}}(\omega_{i}-\omega_{0}) + T_{\mr{r}i} \label{eq:SynGen4}  
	\end{align} 
\end{subequations}
where $\delta_{i}$, $\omega_{i}$, $E'_{i}$, $T_{\mr{M}i}$ denotes the generator's rotor angle, frequency, transient voltage,  and mechanical input torque, respectively, while $E_{\mr{fd}i}$, $T_{\mr{r}i}$ are generator's controllable inputs (exciter field voltage and torque setpoint). 
The constant terms in \eqref{eq:SynGen} are as follows: $M_i$ is the rotor's inertia constant ($\mr{pu} \times \mr{s}^2$), $D_i$ is the damping coefficient ($\mr{pu} \times \mr{s}$),  $x_{di}$ is the direct-axis synchronous reactance ($\mr{pu}$), $x'_{\mr{d}i}$ is the direct-axis transient reactance ($\mr{pu}$), $T'_{\mr{d0}i}$ is the direct-axis open-circuit time constant ($\mr{s}$), $T_{\mr{CH}i}$ is the chest valve time constant, $R_{\mr{D}i}$ is the regulation constant for the speed-governing mechanism, and $\omega_{0}$ denotes the rotor's synchronous speed ($\mathrm{rad/s}$). 
The mathematical model relating generator's internal states $(\delta_{i},\omega_{i},E'_{i},T_{\mr{M}i})$, generator's supplied power $(P_{\mr{G}i},Q_{\mr{G}i})$, and terminal voltage ${v}_i$ is given by the generator's internal algebraic constraint~\cite{sauer2017power}
\begin{subequations}\label{eq:SynGenPower}
	\begin{align}
		\begin{split}
			\hspace{-0.09cm}P_{\mr{G}i} &\hspace{-0.08cm}=\hspace{-0.09cm} \tfrac{1}{x'_{\mr{d}i}}E'_{i}v_i\sin(\delta_i-\theta_i) \hspace{-0.09cm}-\hspace{-0.09cm}\tfrac{x_{\mr{q}i}-x'_{\mr{d}i}}{2x'_{\mr{d}i}x_{\mr{q}i}}v_i^2\sin(2(\delta_i-\theta_i))
		\end{split}
		\label{eq:SynGenPower1} \\
		\begin{split}
			Q_{\mr{G}i} &= \tfrac{1}{x'_{\mr{d}i}}E'_{i}v_i\cos(\delta_i-\theta_i)-\tfrac{x'_{\mr{d}i}+x_{\mr{q}i}}{2x'_{\mr{d}i}x_{\mr{q}i}}v_i^2\\
			\hspace{-0.3cm}&\quad -\tfrac{x_{\mr{q}i}-x'_{\mr{d}i}}{2x'_{\mr{d}i}x_{\mr{q}i}}v_i^2\cos(2(\delta_i-\theta_i)).
		\end{split}\label{eq:SynGePower2}
	\end{align}
\end{subequations}
The power flow equations, for all buses $i \in \mc{N}$, representing the distribution of real and reactive power are given by~\eqref{equ:PF1} and~\eqref{equ:PF2}, which are present in the ACOPF formulation. Hence, the power flow constraints and the generator's algebraic constraints essentially couple the rapidly varying dynamic states and control variables with the ACOPF ones. 

In order to construct the nonlinear state-space representation of the multi-machine power networks~\eqref{equ:PF1}, \eqref{equ:PF2}, \eqref{eq:SynGen}, and \eqref{eq:SynGenPower}, define 
${\m x}_d$ as the vector populating all dynamic states of the network such that
${\m x}_d := \bmat{\m \delta^\top\;\;\m \omega^\top\;\;\m E'^\top\;\;\m T_{\mr{M}}^\top}^\top$ in which ${\m \delta}\hspace{-0.05cm}:=\hspace{-0.05cm}\{\delta_i\}_{i\in \mc{G}}\hspace{-0.05cm}$, ${\m \omega}\hspace{-0.05cm}:=\hspace{-0.05cm}\{\omega_i\}_{i\in \mc{G}}\hspace{-0.05cm}$,  ${\m E'}\hspace{-0.05cm}:=\hspace{-0.05cm}\{E'_i\}_{i\in \mc{G}}\hspace{-0.05cm}$, ${\m T_{\mr{M}}}\hspace{-0.05cm}:=\hspace{-0.05cm}\{T_{\mr{M}i}\}_{i\in \mc{G}}\hspace{-0.05cm}$.  Furthermore, we can define the vector of algebraic states (that overlap with some ACOPF variables) as  $\m x_a := \bmat{\m P_{\mr{G}}^{\top} \;\; \m Q_{\mr{G}}^{\top} \;\;  \m v^\top \;\; \m \theta^\top}^\top$.  The controllable input of the power network is defined as ${\m u} := \bmat{\m E_{\mr{fd}}^\top\;\;\m T_{\mr{r}}^\top}^\top$ where  
${\m E_{\mr{fd}}}\hspace{-0.05cm}:=\hspace{-0.05cm}\{E_{\mr{fd}i}\}_{i\in \mc{G}}\hspace{-0.05cm}$ and ${\m T_{\mr{r}}}\hspace{-0.05cm}:=\hspace{-0.05cm}\{T_{\mr{r}i}\}_{i\in \mc{G}}\hspace{-0.05cm}$. In addition, define the vector ${\m w}$ as ${\m w} :=  \bmat{\m P_{\mr{R}}^{\top} \;\; \m Q_{\mr{R}}^{\top}\;\; \m P_{\mr{L}}^{\top} \;\; \m Q_{\mr{L}}^{\top}}^{\top}$ where ${\m P_{\mr{R}}}\hspace{-0.05cm}:=\hspace{-0.05cm}\{P_{\mr{R}i}\}_{i\in \mc{R}}\hspace{-0.05cm}$, ${\m Q_{\mr{R}}}\hspace{-0.05cm}:=\hspace{-0.05cm}\{Q_{\mr{R}i}\}_{i\in \mc{R}}\hspace{-0.05cm}$, ${\m P_{\mr{L}}}\hspace{-0.05cm}:=\hspace{-0.05cm}\{P_{\mr{L}i}\}_{i\in \mc{L}}\hspace{-0.05cm}$, ${\m Q_{\mr{L}}}\hspace{-0.05cm}:=\hspace{-0.05cm}\{Q_{\mr{L}i}\}_{i\in \mc{L}}\hspace{-0.05cm}$. Essentially, vector $\m w$ lumps all uncertain quantities from renewables and loads. 
The above notations allow us to have a compact, nonlinear differential algebraic equation (NDAE) state space model:
\begin{subequations}\label{eq:nonlinearDAEexplicit}
	\begin{eqnarray}
		\hspace{-0.4cm}\textit{Dynamics:}	& \dot{{\m x}}_d &= {\m A}_d{\m x}_d  +  {\m f}_d\left({\m x}_d,{\m x}_a \right)+  \mB_d\m u \label{eq:nonlinearDAEexplicit-1}\\
		\hspace{-0.4cm}	\textit{Constraints:} &  	\m 0 &= {\m A}_a{\m x}_a + {\m f}_a\left({\m x}_d,{\m x}_a\right) + {\m B}_a {\m w}\label{eq:nonlinearDAEexplicit-2}
	\end{eqnarray}
\end{subequations}
where $\m x_d \in \mbb{R}^{n_d}$, $\m x_a \in \mbb{R}^{n_a}$, $\m u \in \mbb{R}^{n_u}$, and $\m w \in \mbb{R}^{n_w}$. The functions $\m f_d:\mathbb{R}^{n_d}\times \mathbb{R}^{n_a}  \rightarrow \mathbb{R}^{n_{d}}$ and $\m f_a:\mathbb{R}^{n_d}\times \mathbb{R}^{n_a}  \rightarrow \mathbb{R}^{n_{a}}$ defined the vector-valued mapping containing the nonlinearity of generator dynamics as well as the power flow nonlinearity/nonconvexity. Matrices ${\m A}_d\in \mbb{R}^{n_d\times n_d}$, ${\m A}_a\in \mbb{R}^{n_a\times n_a}$, ${\m B}_d\in \mbb{R}^{n_{d}\times n_{d}}$, and ${\m B}_a\in \mbb{R}^{n_{a}\times n_{a}}$ define the linear portion of the dynamics and algebraic constraints. By defining $\m x = \bmat{\m x_d&\m x_a}^\top\in \mbb{R}^{n_x}$ and  $\m f(\m x) = \bmat{\m f_d(\m x_d, \m x_a) & \m f_a(\m x_d, \m x_a)}^\top$ the model \eqref{eq:nonlinearDAEexplicit} can also be rewritten as follows:
\begin{align}\label{eq:final_NDAE_cntrl}
	\m E\dot{{\m x}} &= {\m A}{\m x} +  {\m f}\left({\m x} \right) + {\m B} {\m u } + {\m B}_w \m w.
\end{align}
Having defined the NDAE power network dynamics, we note the following. \textit{(i)} Herein, we showcase a fourth-order generator model (i.e., each generator is modeled via four states) but this can be extended to higher-order generator dynamics as well as dynamic models of solar and wind. \textit{(ii)} In addition to modeling the algebraic constraints encoding lossy power flows, the presented NDAE formulation also accounts for the stator algebraic equation which is usually missing from ACOPF formulation. \textit{(iii)} The controllable variable in the ACOPF formulation, namely $\m P_{\mr{G}}$, is present in the dynamical system model as an algebraic variable that is controlled explicitly via $\m u(t)$. This entails the following. Solving a feedback control problem that generates realtime sequence  $\m u(t)$ and subsequently extracting the ACOPF's algebraic variables $\m x_a(t)$, while satisfying the ACOPF constraints and being close to its optimal solution $J_{\mr{OPF}}(\m P_\mr{G})$, could be specifically useful. 

\section{Solving OPF via DAE Control Theory}\label{sec:NDAE_control}

We focus now on the control problem for the NDAE model \eqref{eq:nonlinearDAEexplicit}, which when solved will essentially solve a version of the ACOPF~\eqref{equ:ACOPF}. This control problem can simply be defined as computing a constant gain matrix that can be used with the control input $\m u(t)$ in a closed-loop fashion (via realtime state/output information) such that it can drive the system back to a stable equilibrium after a large disturbance. With that in mind, let us
define the closed-loop system dynamics as follows:
\begin{align}\label{eq:final_NDAE_cntrl_closed}
	\m E\dot{{\m x}} &= {\m A}{\m x} +  {\m f}\left({\m x} \right) + {\m B} {\m u_C } + {\m B}_w \m w
\end{align}
where $\m u_C$ is the closed-loop control input and is defined as:
\begin{align}\label{eq:ucl}
	\boxed{{\m u}_{{C}}:= \m u_{{C}}(t) = \m u_{\mathrm{ref}}^k + \m K \left({\m x}(t) - \m x^k\right)}
\end{align}
in which $\m u_\mr{ref}$ is the reference or baseline setting for the control input $\m u$, $\m x^k$ is the dynamic states information at previous time step $k$, and $\mK$ is the constant controller gain matrix. Notice that, $\m u_\mr{ref}$ and $\m x^k$ can be determined numerically using power flow studies. That being said, the key idea is to design $\mK$ such that using realtime state feedback information $\m x(t)$, the closed-loop control input $\m u_{C}$ can make the system robust and transiently stable against disturbances.  

To that end, notice that, if we can compute $\mK$ in a way such that it encodes \eqref{eq:nonlinearDAEexplicit-2} also along with \eqref{eq:nonlinearDAEexplicit-1} (meaning determining $\mK$ for the whole NDAE system instead of eliminating \eqref{eq:nonlinearDAEexplicit-2} and converting it to an ODE system), then the determined feedback controller $\mK$ will inherently satisfy the key constraints appearing in the OPF formulation \eqref{equ:ACOPF}. This is because Eq. \eqref{eq:nonlinearDAEexplicit-2} includes power balance equations \eqref{equ:PF1}, \eqref{equ:PF2} of ACOPF and generators stator algebraic constraints \eqref{eq:SynGenPower1},\eqref{eq:SynGePower2} which indirectly encode the constraints \eqref{equ:PG_constratints},\eqref{equ:QG_constraints} of the ACOPF. As for the other constraints such as limits on generators' capacities, these can be encoded via saturation dynamics in the differential equations. Admittedly, other constraints such as thermal limits of lines cannot be modeled in this approach, and to that end we thoroughly investigate any constrained violations incurred in Section~\ref{sec:casestudies}. 

With that in mind, we name the computation of such feedback controller gain $\mK$ which includes \eqref{eq:nonlinearDAEexplicit-2} in its control architecture as \textit{control-OPF} feedback controller design. This is because such  $\mK$ ensures system transient stability after a large disturbance and also fully abides by the key OPF constraints as discussed above. To that end, we present the following results to compute such $\mK$ which is based on Lyapunov stability theory as follows:
\begin{align*}
	\small{\textsc{(\textbf{Control-OPF})}}\minimize_{\m F,\m S,\m X, {\lambda},\epsilon,\mu,\kappa} &\;\;\; a_1\lambda\hspace{-0.01cm}+\hspace{-0.01cm}a_2\mu\hspace{-0.01cm}+\hspace{-0.01cm}a_3\epsilon_1\hspace{-0.01cm}+\hspace{-0.01cm}a_4\epsilon_2\\ \subjectto & \;\;\;\mr{LMI}\; \eqref{eq:LMI_Hinf},\; \mr{LMIs}\; \eqref{eq:LMIs_size},\\&\;\;\; \lambda \mI - \mX^\top\mE\mX \succ 0,\\& \;\;\;\m X \succ 0,\lambda>0,\gamma>0,\\&\;\;\;\epsilon_1>0,\epsilon_2>0, \kappa>0
	\vspace{-0.6cm}
\end{align*}
where
\vspace{-0.1cm}
\begin{itemize}
	\item $a_1$, $a_2$, $a_3$, and $a_4$ are known weighting constants.
	\item The variables in control-OPF are matrices $\mS\in \mbb{R}^{n_a\times n_x}$, $\mF\in \mbb{R}^{n_u\times n_x}$, $\m X \in \mbb{S}_{++}^{n_x \times n_x}$, and scalars $\lambda, \mu,\kappa, \epsilon_1, \epsilon_2 \in\mbb{R}_{++}$.
	\item LMI \eqref{eq:LMI_Hinf} is defined as
	\begin{align}\label{eq:LMI_Hinf}
		\hspace{-0.4cm}\bmat{\mT^\top\hspace{-0.09cm}\mA^\top_{c}\hspace{-0.09cm}+\hspace{-0.09cm}\mA_{c}\mT & \mB_w&\mI &\mT^\top\hspace{-0.09cm}\m\Phi^\top  & \kappa^\frac{1}{2}\alpha\mT^\top \\ \mB_w^\top & -\mu\mI &\mO& \mD_w^\top &\mO \\ \mI& \mO& -\kappa\mI&\mO&\mO\\ \m\Phi\mT&\mD_w&\mO&-\mI&\mO\\ \kappa^\frac{1}{2}\alpha\mT& \mO& \mO& \mO& -\mI}	\hspace{-0.09cm} \prec	\hspace{-0.001cm} 0
	\end{align}
	where $\mT = \mX\mE^\top+\mE^\perp\mS$ and $\mE^{\perp}\in\mbb{R}^{n_x\times n_a}$ represent the orthogonal complement of matrix $\mE$.
	\item LMIs \eqref{eq:LMIs_size} are defined as:
	\begin{align}\label{eq:LMIs_size}
		\bmat{-\epsilon_1\mI&\mF^\top\\\mF&-\mI}\prec0, \,\,\,\,\, \bmat{-\epsilon_2\mI&\mI\\\mI&-\mT-\mT^\top}\prec 0
	\end{align}
	\item The corresponding controller gain matrix $\m K$ can be obtained as
	\begin{align}\label{eq: gain K}
		\m K = \mF\mT^{-1}
	\end{align}
	\item The designed control-OPF is a convex semi-definite optimization problem and thus can easily be solved via various optimization solvers. 
\end{itemize}
It is worth mentioning here that the designed control law $\m u_C(t)$ of the proposed control-OPF requires that all the state variables (both $\m x_d$ and $\m x_a$) are known in realtime. This requirement can easily be satisfied these days because of the recent developments in synchronized measurement technologies and highly efficient state estimation algorithms \cite{Liu2021TPWRS, ZhaoTPWRS2019, JunjianITSG2018, ZhaoITSG2019, ZhouITPWRS2013}.  These modern state estimation algorithms only require measurements from a few PMUs (placed optimally such that the system is observable) and can efficiently estimate all the states of the network including the states of solar plants \cite{NadeemCDC2022, NadeemITPWRS2022}.

In the following sequel, we present a detailed explanation of each variable and the mathematical derivation of the proposed control-OPF formulation.
\subsection{Mathematical Derivation of Control-OPF}
To begin with, let us assume there is an unknown disturbance in the system and the new value of the vector $\m w$ is $\m w'$. This disturbance will move the states of the power network from its initial equilibrium to a new equilibrium $\m x'$. With that in mind, the perturbed closed loop dynamics i.e $\Delta \m x = \m x - \m x'$ of \eqref{eq:final_NDAE_cntrl_closed} can be written as:
\begin{align}\label{eq:final_NDAE_cntrl_closed_per}
	\m E\Delta\dot{\m x}\hspace{-0.05cm} &=\hspace{-0.05cm} (\m A\hspace{-0.05cm}+\hspace{-0.05cm}\m{BK})\Delta\m x+\Delta\m f(\Delta \m x)\hspace{-0.05cm} + \hspace{-0.05cm}\m B_w \Delta\m w.
\end{align}
Now the main idea is to design in $\m K$ such that the perturbed dynamics \eqref{eq:final_NDAE_cntrl_closed_per} converge asymptotically to zero. With that in mind, to compute such controller gain matrix $\mK$ we utilize the robust $\mathcal{H}_\infty$ notion \cite{zhou1998essentials}. The core idea in  $\mathcal{H}_\infty$-based controller design is that the controller makes sure that the norm of the performance index is always less than constant times the norm of disturbance i.e $\norm{\m z_1}^2_{\mathcal{L}_2} <  \gamma^2\norm{{\m w }}^2_{\mathcal{L}_2}$, where $\m z$ is the user-defined performance index and $\gamma$ is optimization variable commonly known as performance level in the control theocratic literature. To that end, let us consider the performance index for the perturbed closed loop system \eqref{eq:final_NDAE_cntrl_closed_per} to be: $\Delta \m z_1= \m C \Delta\m x + \mD\Delta\m u  + \mD_w\Delta\m w \in \mbb{R}^{n_x}$ which can also be written as: $\Delta \m z_1= (\m C+\m{DK})\Delta\m x + \mD_w\Delta\m w$, where the matrices $\m C \in \mbb{R}^{n_x \times n_x}$, $\m D \in \mbb{R}^{n_x \times n_u}$, and $\m D_w \in \mbb{R}^{n_x \times n_w}$ are user-defined penalizing matrices (meaning how much weight should be given to each state and control inputs in response to the disturbance) similar to the $\mQ$, $\m R$ matrices in LQR type controller. From now on for the sake of notation simplicity with little abuse of notation, we will consider $\Delta\m x = \m x$, $\Delta\m w = \m w$, and $\Delta\m z = \m z$. 

That being said, we now utilize Lyapunov stability theory to design the controller gain $\mK$ which guarantees $\mathcal{H}_\infty$ stability of the closed loop system \eqref{eq:final_NDAE_cntrl_closed_per}. To that end, let us consider a Lyapunov function $V(\m x)=\m x^\top \mE^\top \mP \m x $, with $\m P\in\mbb{R}^{n_x\times n_x}$ and $V:\mbb{R}^{n_x}\rightarrow \mbb{R}_+$. Now assuming that the well-known Kalman-Popov-Yakubovich (KYP) lemma \cite{boyd1994linear} $\mE^\top \mP = \mP^\top \mE \succeq \mO$ satisfies, then the derivative of $V$ along system trajectories $\m x$ can be written as follows:
\begin{align*}
	{\dot V}(\m x) &= (\m{E}\dot{\m x})^\top \mP \m x+(\mP \m x)^\top(\m{E}\dot{\m x}).
\end{align*}
Now $\mathcal{H}_\infty$ criterion can be written as ${\dot V}(\m x) + \m z_1^\top\m z_1 - \gamma^2{\m w}^\top{\m w} < 0$, then
by plugging the value of $\m E\dot{\m x}$ from \eqref{eq:final_NDAE_cntrl_closed_per} in it and doing simplifications we get the following quadratic form $\m\Psi^\top\m\Upsilon\m\Psi<0$, with $\m\Psi\hspace{-0.01cm} = \hspace{-0.03cm}\bmat{\m x& \m w & \Delta\m f}^\top$ and
\begin{align*}
	\m\Upsilon=\hspace{-0.1cm}\bmat{ \mA^\top_{c}\mP+\m\Phi^\top\m\Phi+\mP^\top \mA_{c}& \m\Phi^\top{\m D}_w + \mP^\top{\m B}_w  & \mP^\top\\  {\m D}_w^\top\m\Phi + {\m B}_w^\top\mP& {\m D}_w^\top{\m D}_w-\gamma^2\mI&\mO\\\mP& \mO&\mO}
\end{align*}
where $\mA_c =\mA+\m{BK}$ and $\m\Phi =\mC+\m{DK}$. Notice  $\m\Psi^\top\m\Upsilon\m\Psi<0$ holds if  $\m\Upsilon\prec0$. Now assuming $\Delta\m f$ to be quadratically-bounded with known constant $\alpha$ such that
\begin{align}
	\norm{\Delta{\m f(\m x) }}_2 \leq \norm {{\alpha( \m x)}}_2 \label{eq:13}\\
	\Leftrightarrow {\Delta\m f(\m x)}^\top  {\Delta\m f(\m x)}-\alpha^2\m x^\top\m x\leq 0
\end{align}
which can be written as $\m\Psi^\top\m \Theta\m\Psi\leq0$, where
\begin{align*}
	\m \Theta = \mr{diag}\left( \bmat{-\alpha^2&\mO&\mI}\right). 
\end{align*}
Notice that it is common in the power system feedback control literature to assume some sort of boundedness (such as norm or quadratic boundedness or Lipschitz continuity assumption) on the structure of the nonlinearity to design feedback controllers in a tractable fashion \cite{nadeem2023robust, siljak2002robust, nadeem2022dynamic}. These assumptions are applicable to power systems models since the system states have specific upper and lower limits. For instance, the voltage is constrained to lie within a range from $0.95$ pu to $1.05$ pu. That being said, we proceed with the derivation of the control-OPF formulation. By S-Lemma \cite{Slemma}, if there exists a scalar $\kappa \geq 0 $ then  $\m\Upsilon-(\kappa)\m \Theta \prec0$ holds, which can be written as:
\begin{align*}
	\hspace{-0.1cm}\bmat{ \mA^\top_{c}\mP\hspace{-0.05cm}+\hspace{-0.05cm}\m\Phi^\top\m\Phi\hspace{-0.05cm}+\hspace{-0.05cm}\mP^\top \mA_{c}\hspace{-0.05cm}+\hspace{-0.05cm}\kappa\alpha^2\mI& \m\Phi^\top{\m D}_w\hspace{-0.05cm} +\hspace{-0.05cm} \mP^\top{\m B}_w  & \mP^\top\\  {\m D}_w^\top\m\Phi \hspace{-0.03cm}+\hspace{-0.03cm} {\m B}_w^\top\mP& {\m D}_w^\top{\m D}_w\hspace{-0.03cm}-\hspace{-0.03cm}\gamma^2\mI&\mO\\\mP& \mO&-\kappa\mI}.
\end{align*}
Now applying congruence transformation with $\mr{diag}\left( \bmat{\m T^\top&\mI&\mI}\right)$ where $\mT = \mP^{-1}$, then the above matrix inequality can be written as follows:
\begin{align}\label{eq:prf_cong}
	\hspace{-0.1cm}\bmat{\m\Omega & \mT^\top\m\Phi^\top{\m D}_w\hspace{-0.03cm} +\hspace{-0.03cm}{\m B}_w  & \mI\\  {\m D}_w^\top\m\Phi\mT \hspace{-0.03cm}+\hspace{-0.03cm} {\m B}_w^\top& {\m D}_w^\top{\m D}_w\hspace{-0.03cm}-\hspace{-0.03cm}\gamma^2\mI&\mO\\\mI& \mO&-\kappa\mI}
\end{align}
with 
$\m\Omega=\mT^\top\mA^\top_{c}\hspace{-0.03cm}+\hspace{-0.03cm}\mT^\top\m\Phi^\top\m\Phi\mT\hspace{-0.03cm}+\hspace{-0.03cm} \mA_{c}\mT\hspace{-0.03cm}+\hspace{-0.03cm}\kappa\alpha^2\mT^\top\mT$. Using Schur complement lemma \cite{zhang2006schur} on \eqref{eq:prf_cong} we get
\begin{align}\label{eq: prf 14}
	\hspace{-0.1cm}\bmat{\mT^\top\mA^\top_{c}\hspace{-0.03cm}+\hspace{-0.03cm}\mA_{c}\mT & \mB_w&\mI &\mT^\top\m\Phi^\top  & \kappa^\frac{1}{2}\alpha\mT^\top \\ \mB_w^\top & -\gamma^2\mI &\mO& \mD_w^\top &\mO \\ \mI& \mO& -\kappa\mI&\mO&\mO\\ \m\Phi\mT&\mD_w&\mO&-\mI&\mO\\ \kappa^\frac{1}{2}\alpha\mT& \mO& \mO& \mO& -\mI}
\end{align}
Now to get a strict LMI for controller design we have to eliminate the KYP lemma, this can be done as detailed in \cite{nadeem2023robust, nadeem2023wide}. That being said, let us assume there exist matrices 
${\m U}\in \mbb{R}^{n_x\times n_x}$ and ${\m V}\in \mbb{R}^{n_x\times n_x}$ such that
\begin{align}\label{eq:proof-eq-1}
	\begin{split}
		\m{U E V} &= \bmat{\m I & \m O\\ \mO & \mO}, 
	\end{split}
	\begin{split}
		{(\mU^{-1})^\top \mP \mV} &= \bmat{\mP_1 & \mP_2\\ \mP_3 & \mP_4}. 
	\end{split}
\end{align}
Then from \eqref{eq:proof-eq-1} we get 
\begin{subequations}\label{eq:proof-eq-3}
	\begin{align}
		\begin{split}\label{eq:proof-eq-3a}
			\mE^\top \mP = (\mV^{-1})^\top\bmat{\mP_1&\mO\\\mO&\mO}\mV^{-1}
		\end{split}\\
		\begin{split}\label{eq:proof-eq-3b}
			\mP^\top \mE = (\mV^{-1})^\top\bmat{\mP_1^\top&\mO\\\mP_2^\top&\mO}\mV^{-1}.
		\end{split}
	\end{align}
\end{subequations}
Then we can see from \eqref{eq:proof-eq-3} that $\mP^\top \mE$ and $\mE^\top \mP$  can be made equal only if $\m{P_2}^\top = \mO$ and $\m{P_1} = \m{P_1}^\top$. Hence, $\mP$ can be updated as
\begin{align*}
	\begin{split}
		\m{P} &\hspace{-0.0cm}= \hspace{-0.0cm}\mU^\top\underbrace{\bmat{\mP_1 & \m {O}\\ \mP_3 & \mP_4}}_{\m{\bar P}}\mV^{-1}.
	\end{split}
\end{align*}
and $\mT = \mP^{-1}$ can be written as
\begin{align*}
	\begin{split}
		\m{T} &\hspace{-0.0cm}= \hspace{-0.0cm}\mV\underbrace{\bmat{\mP^{'}_1 & \m {O}\\ \mP^{'}_3 & \mP^{'}_4}}_{\m{\bar P}^{-1}}(\mU^{-1})^\top
	\end{split}
\end{align*}
Then, for any $\m X \in \mbb{S}_{++}^{n_x \times n_x}$ it is straightforward to show 
\begin{align}\label{eq:proof-eq-finalb}
	\m{T} &= \m{XE}^\top+\mE^{\perp}{\mS}
\end{align}
\noindent where $\mS\in \mbb{R}^{n_a\times n_x}$ and $\mE^{\perp}\in\mbb{R}^{n_x\times n_a}$ is the orthogonal  complement of matrix $\mE$ . Finally, by defining $\mF = \m{KT}$, $\mu=\gamma^2$, and plugging the value of $\mT$ from \eqref{eq:proof-eq-finalb} into \eqref{eq: prf 14} we get the LMI \eqref{eq:LMI_Hinf}. Notice that one can minimize the maximum eigenvalues of the assumed candidate Lyapunov function to ensure quick convergence, which can be written as $\lambda \mI - \mX^\top\mE\mX \succ 0$ with $\lambda$ being an optimization variable that should be minimized as shown in the proposed control-OPF formulation. Furthermore, as  $\mK = \mF\mT^{-1}$ then one can limit the size of controller gain $\mK$ by constraining $\mF$ and $\mT$ as: $\mH^\top\mH\prec \epsilon_1\mI$ and $\mT^{-1}\prec \epsilon_2\mI$. Which in LMI formulation (via applying the Schur complement) can equivalently be written as LMIs \eqref{eq:LMIs_size}. This ends the derivation of control-OPF.

By solving control-OPF we can determine an appropriate time-invariant gain matrix $\m K$ as given in \eqref{eq: gain K} which can be plugged into \eqref{eq:ucl} to design a feedback control law that guarantees the stability of the system after a large disturbance. Notice that the computation of $\mK$ is carried out offline.\footnote{This matrix gain is computed offline as it does \textit{not} depend on the state of the system and only relies on the system's parameters and topology. Hence, its computation is performed offline. In case topological changes happen in the system, this gain matrix $\m K$ should ideally be recomputed, but feedback control gains are known to be robust to minor changes in system parameters and topology.}  Furthermore, the design control law $\m u_{C}$ acts in realtime based on the system state/output information provided by PMUs in power systems. Notice that, the overall proposed controller design in this work is different than 
\cite{SinghTPWRS2016, AranyaICSM2019, nadeem2023wide, nugroho2023load}, as here we are utilizing robust $\mathcal{H}_\infty$ notion, considering nonlinearities also in controller design, and also engineering the overall controller architecture as an efficient optimization problem which ensures quick convergence of state variables with optimal controller gain matrix $\mK$.

\subsection{Control-OPF Novelty, Properties, and  Literature Discussion}

We want to emphasize here that the control-OPF gain matrix $\mK$ has been derived in a way such that it satisfies the algebraic constraints \eqref{eq:nonlinearDAEexplicit-2} of the NDAE power system model. This can also be verified by looking at the structure of the proposed LMI \eqref{eq:LMI_Hinf}, we can observe that it is dependent on the singular matrix $\mE$ and the whole system matrices $\mA$, $\mB$ and $\mB_w$ (which encodes the algebraic constraints matrices $\mA_a$ $\mB_a$) of Eq. \eqref{eq:nonlinearDAEexplicit-2}. This means that  $\mK$ inherently satisfies some of the key constraints appearing in ACOPF formulation \eqref{equ:ACOPF}. Although the rest of the ACOPF constraints such as line thermal limits and voltage limits are not explicitly modeled in the presented control-OPF architecture, through extensive numerical case studies under various conditions we show that these constraints are also indeed satisfied. This is because the control-OPF also makes sure that the system is transiently stable after a large disturbance. 

Furthermore, notice that matrix $\mD$ is a penalizing matrix on the control inputs, meaning how much control effort needs to be performed by each generator in response to the disturbance. In this way, we can control how much active power needs to be extracted from a particular generator to meet the varying load demand. Thus by appropriately designing $\mD$ (i.e., by putting more penalty on those generators that are expensive), one can ensure that after a large disturbance, the system operating cost is optimal. In this work, as the control inputs are field voltage $\mE_{\mr{fd}}$ and governor reference valve position $\mT_r$ and since $\mT_r$ directly control active power output from the generators, then in designing $\mD$ matrix a larger penalty has been added to the $\mT_r$ of those generators which are expensive (by looking at the quadratic cost function of each generator).

To that end, since the control-OPF acts in realtime and provides stability guarantees while also satisfying ACOPF conditions then the need for running ACOPF after $5-10$ minutes in the tertiary layer of the power system can be eliminated. Thus, we essentially dumped the ACOPF problem in a feedback control architecture.  It is worthwhile to mention that in the presented control-OPF we do not even need to solve the power system NDAE model. In a real-world application, the NDAE \eqref{eq:nonlinearDAEexplicit} is replaced by the actual power system model. Thus the control-OPF is essentially carried out offline and then $\mK$ is used online, knowing that $\mK$ satisfies system algebraic constraints.


It is worth mentioning here that, as compared to the literature where authors have tried to merge system optimality and secondary control (as in \cite{li2015connecting, miao2016achieving, eidson1995advanced, zhao2018distributed, mallada2017optimal, zhao2015distributed}) this work differs in the following ways. Here, in response to the system transients, the controller adjusts the power output of all the generators optimally (via properly setting design matrices $\m C$ and $\m D$) through realtime information received from the PMUs while also satisfying the algebraic constraints. Furthermore, the proposed controller in this work can directly be actuated through the primary layers controllers, for example in the case of $4^{th}$-order generator model as used in this paper, the proposed controller is actuated through turbine dynamics (via torque setpoint $\m T_r$) and by directly controlling the generator field voltage $\m E_{fd}$. In the case of higher-order generator dynamics (which include automatic voltage regulator (AVRs) dynamics) the proposed controller can adjust the setpoint of AVRs and turbine dynamics.

Furthermore, the studies \cite{zhao2018distributed,mallada2017optimal} design cost-optimal frequency controller; however, the optimality of the controller is not clear as no-cost comparison with OPF has been carried out.  In addition, the designed controller requires controllable loads to improve system performance, and without controllable loads, the proposed controller works exactly the same as AGC. Similarly in \cite{li2015connecting, miao2016achieving, eidson1995advanced} the
economic dispatch layer (or the OPF layer) has been merged with the AGC layer via formulating an optimization problem that solves them simultaneously, however the design methodology cannot improve system transient stability (by adding damping to the system oscillations) and can only remove steady-state error in the frequency (similar to the AGC but optimally). These methodologies are not capable of incorporating generator dynamics and realtime PMUs data in their proposed methodologies.

We also want to point out here that the proposed approach in this work is based on linear matrix inequalities and is hence convex. However, it is completely different than the methods in the literature that solve ACOPF using convex relaxation or linearization. This is because to find a solution for ACOPF problem convex relaxation or linearization is commonly carried out for the highly nonlinear nonconvex AC power flow equations \cite{gopalakrishnan2012global,lu2018tight, LeeITPWRS2020}. While on the other hand, the proposed approach does not carry out any convex relaxation or linearization and directly encodes the algebraic constraint model (which models these nonlinear power flow equations) in its design. Furthermore, notice that the conventional ACOPF does not consider the load/renewable uncertainty and assumes that the forecasted operational conditions used in the OPF formulations are exactly the same as the actual conditions, which is unrealistic. This is mainly because of the increasing penetration of stochastic renewable resources, often observed as substantial fluctuations in load demand caused by behind-the-meter PV power plants. Thus, resulting in significant deviations between the actual operating conditions and the original forecasted conditions used in the OPF formulations. These forecast errors can have consequences, potentially leading to violations of critical operational limits and jeopardizing the system's steady-state stability \cite{lee2021robust}. In this regard, recently many robust AC-OPF formulations have also been proposed---see \cite{molzahn2018towards, roald2013analytical, bienstock2014chance, venzke2017convex}. However, to solve the power flow equation most of these studies again use linearization and convex relaxations. This is because guaranteeing a solution to the AC power flow equation is particularly challenging because of the highly nonlinear relationship between the decision variables. Satisfying the power balance equations is important because it is a necessary condition for system stability.

Furthermore, most of the literature in robust OPF considers a deterministic uncertainty, meaning disturbance realization/set is considered to be \textit{known}, which is unrealistic. Such as in \cite{lee2021robust} robust OPF is proposed with uncertainty in load demand to be in a known deterministic set.  This uncertainty in load demand is first modeled in the power balance equation and then various relaxations and restrictions are carried out to reach the final convex robust AC-OPF formulation. Also, albeit convex robust OPF can give feasible solutions (meaning the solutions satisfy the power balance equations and the operational constraints) against uncertainty in power injections and load demand, they cannot make the system transiently stable after large disturbance as they are control unaware. On the other hand, the proposed control-OPF is truly uncertainty-unaware and can also make the system transiently stable by providing damping to the system oscillations and bringing the system back to its equilibrium after a large disturbance.

\section{Numerical Case Studies}\label{sec:casestudies}
\begin{figure}[]
	\hspace{-0.0cm}\subfloat{\includegraphics[keepaspectratio=true,scale=0.51]{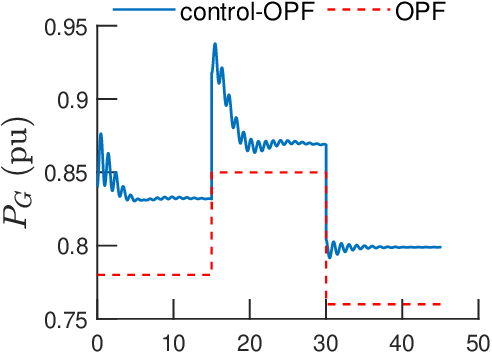}}{}{}\hspace{-0.5cm}
	\hspace{-0.cm}\subfloat{\includegraphics[keepaspectratio=true,scale=0.51]{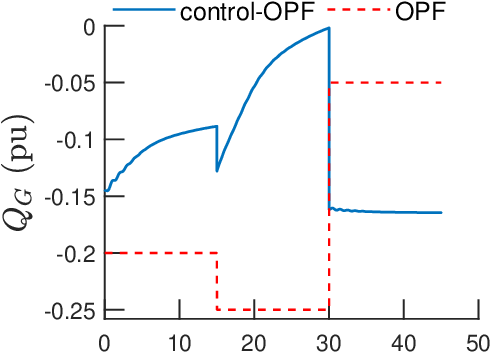}}{}{}\vspace{0.1cm}
	\hspace{-0.0cm}\subfloat{\includegraphics[keepaspectratio=true,scale=0.51]{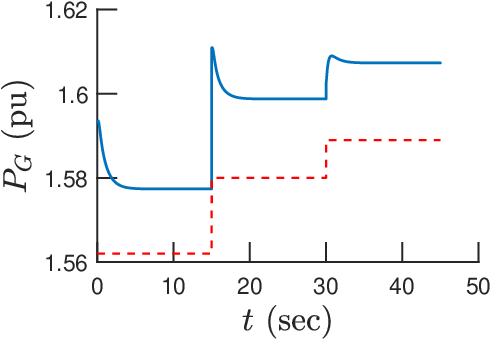}}{}{}\hspace{-0.0cm}
	\subfloat{\includegraphics[keepaspectratio=true,scale=0.51]{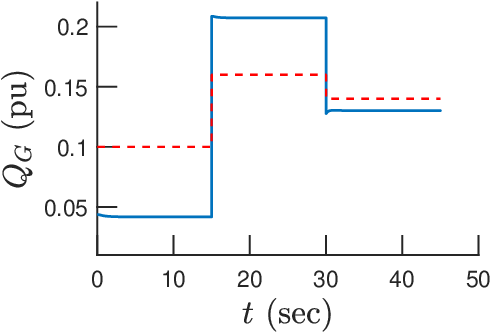}}{}{}\vspace{-0.3cm}
	\caption{Time-varying active/reactive power set-points provided by control-OPF and static set-points from ACOPF for three random step disturbances in load demand; above figures are for case 39 and below figures are for case 9 test system.}\label{fig:Pg Qg all caseA}\vspace{-0.1cm}
\end{figure}

To evaluate the performance of the proposed methodology, we test various magnitudes of disturbances in load and renewable energy resources. We also compare the overall cost of the system with the control-OPF and by just running ACOPF. Notice that the control-OPF provides us time-varying vectors of $\mP_G$ and $\mQ_G$ as shown in Fig. \ref{fig:Pg Qg all caseA} while ACOPF gives static set-points for the generator power outputs. This is because when a disturbance is applied to the system the control-OPF also commands all the generators to increase or decrease power in order to mitigate the effect of the disturbance on the system dynamics. To compute the cost of the system with control-OPF, we evaluate the quadratic cost equation of the generator for the vector  $\mP_G$ (generated from running the control law $\m u(t)$), and then computing the mean of the total cost vector, given as follows:
\begin{align*}
	J_{\mr{OPF}}(\m P_\mr{G}^{\text{control-OPF}}) = \dfrac{1}{T}\sum^T_{t =1} \sum_{i \in \mathcal{G}}a_i P^2_{\mr{G}i}(t) + b_i P_{\mr{G}i}(t)  + c_i  
\end{align*}

\begin{figure}[t]
	\centering
	\subfloat{\includegraphics[keepaspectratio=true,scale=0.6]{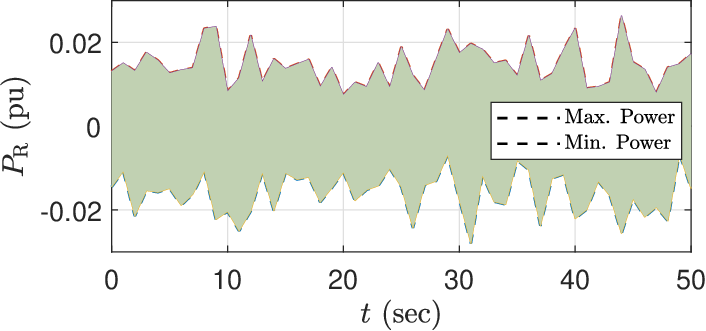}}{}{}\hspace{-0.1cm}
	\caption{Random uncertainty in renewable power generations}\label{fig:random_renew}
\end{figure}

With that in mind, two case studies are carried out as discussed in the below sections. In the first case study, we apply random step uncertainty in load demand with high Gaussian noise and evaluate the system total cost and compare it with ACOPF cost. A similar comparison has been carried out in the second case study. However, here we also assume high uncertainty in the power generated by renewables as shown in Fig. \ref{fig:random_renew}. In both case studies, we also check if with the control-OPF the system violates any ACOPF constraints or not.

To that end, in this section, the following high-level research questions are investigated.
\begin{itemize}
	\item \textit{Q1.} Given that the control-OPF strategy does not explicitly take into account generators' cost curves and the ACOPF cost function $J_{\mr{OPF}}(\m P_\mr{G}^{\text{ACOPF}})$, how far are the generators' varying setpoints and their corresponding aggregate costs $J_{\mr{OPF}}(\m P_\mr{G}^{\text{control-OPF}})$ from the ACOPF solutions?
	\item \textit{Q2.} Can we quantify the price of realtime control and regulation of the grid's dynamic states? 
	\item \textit{Q3.} The control-OPF approach does not take into account inequality constraints modeling thermal line limits. Does this approach result in any constraint violations of the ACOPF? 
	\item \textit{Q4.} Is the comparison between ACOPF and control-OPF fair? While the former know exact values for all uncertain loads and renewables (needed to compute ACOPF setpoints), the latter is truly uncertainty-unaware. 
\end{itemize}
\begin{figure}
	\centering
	\includegraphics[keepaspectratio,scale=0.450]{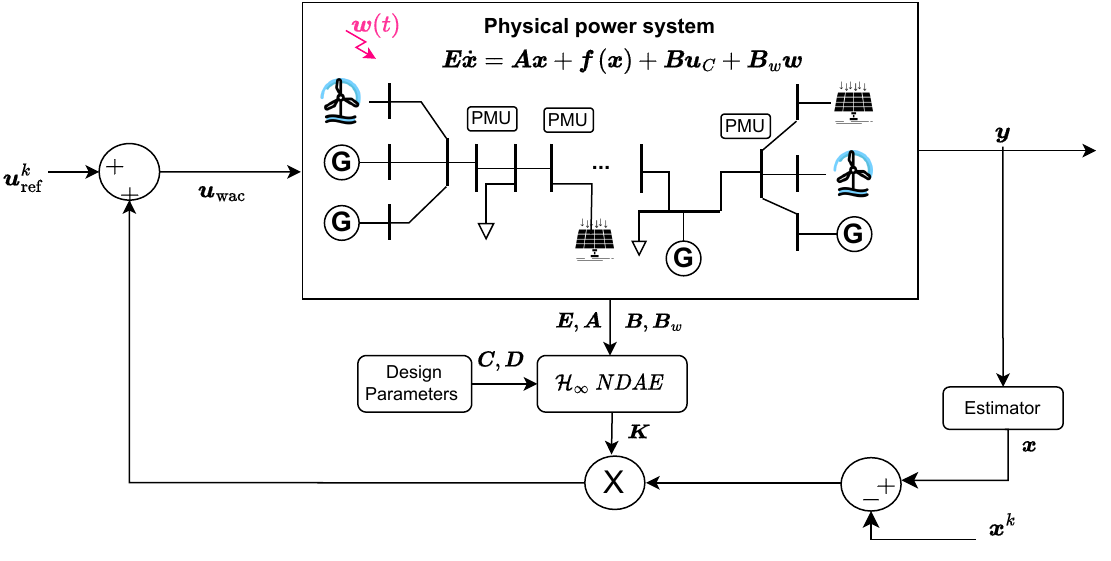}\vspace{-0.0cm}\caption{Overall integrated framework of the proposed control-OPF}\label{fig:design}\vspace{-0.0cm}
\end{figure}
All the simulation studies are carried out in MATLAB $2022b$ and using MATPOWER software. Optimal power flow for all the case studies is carried out by running  \texttt{runopf} command in MATPOWER \cite{MATPOWER2011}. The control-OPF gain is computed via YALMIP \cite{Lofberg2004} and using MOSEK \cite{Andersen2000} solver, while the power system NDAEs \eqref{eq:nonlinearDAEexplicit} are simulated using MATLAB DAEs solver \texttt{ODE15i}. The overall architecture of the control-OPF can be seen in Fig. \ref{fig:design}.

\begin{figure}[t]
	\hspace{-0.0cm}\subfloat{\includegraphics[keepaspectratio=true,scale=0.51]{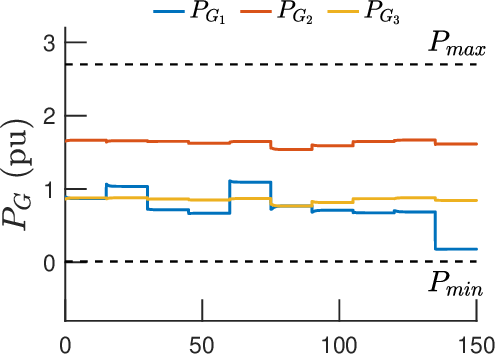}}{}{}\hspace{-0.15cm}
	\subfloat{\includegraphics[keepaspectratio=true,scale=0.51]{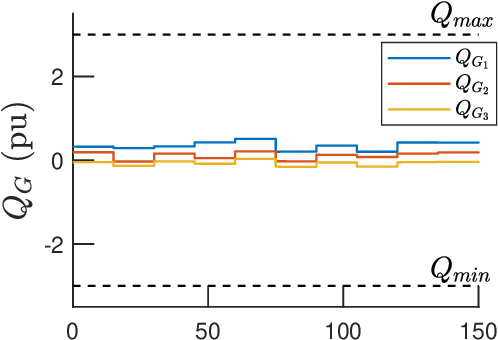}}{}{}\vspace{0.19cm}
	\subfloat{\includegraphics[keepaspectratio=true,scale=0.51]{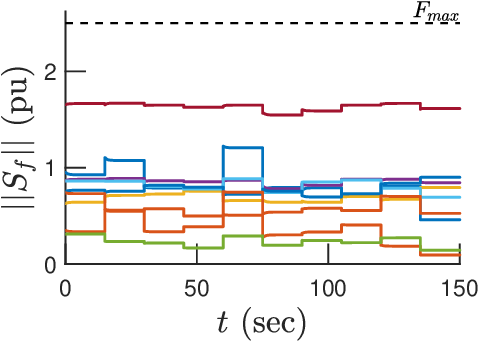}}{}{}\hspace{-0.1cm}
	\subfloat{\includegraphics[keepaspectratio=true,scale=0.48]{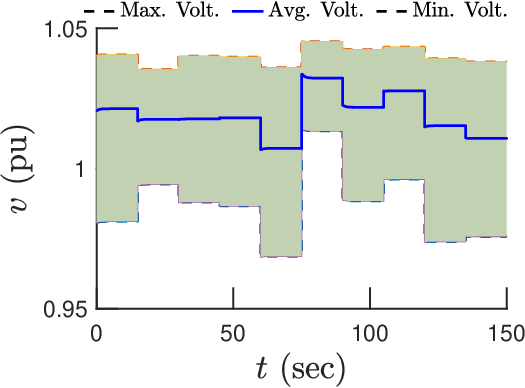}}{}{}\vspace{0.19cm}
	\caption{Active and reactive power generated by all the generators and their respective limits, line flows and their maximum rating, and the overall modulus of all bus voltages for case 9 bus test system for Scenario A.}\label{fig:const_case9_caseA}
	\vspace{-0.5cm}
\end{figure}

\begin{figure}[t]
	\hspace{-0.2cm}\subfloat{\includegraphics[keepaspectratio=true,scale=0.51]{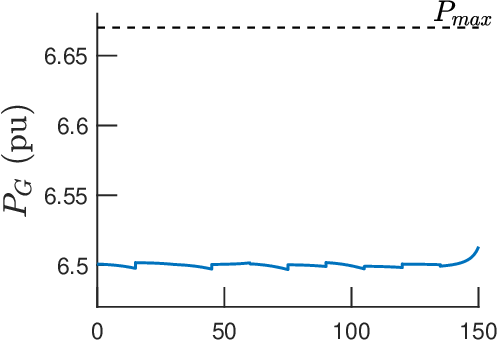}}{}{}\hspace{-0.15cm}
	\subfloat{\includegraphics[keepaspectratio=true,scale=0.51]{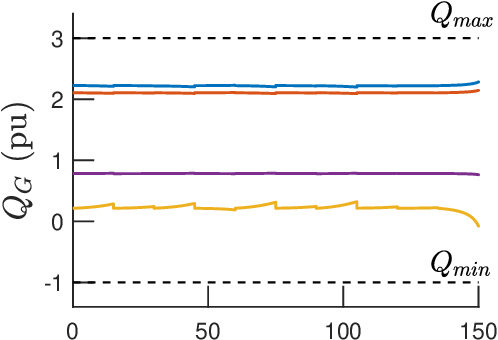}}{}{}\vspace{0.0cm}
	\hspace{-0.2cm}\subfloat{\includegraphics[keepaspectratio=true,scale=0.51]{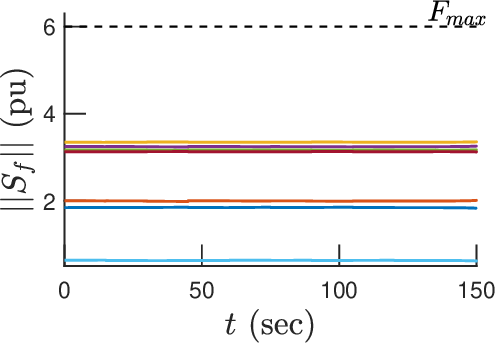}}{}{}\hspace{-0.13cm}
	\subfloat{\includegraphics[keepaspectratio=true,scale=0.48]{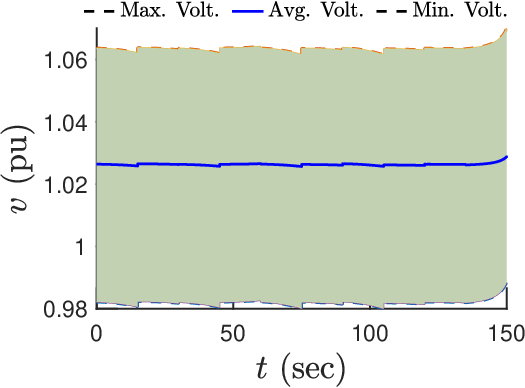}}{}{}\vspace{0.15cm}
	\caption{Active and reactive power of a couple of generators and their respective limits, line flows, and their maximum rating, and the overall modulus of all buses voltages for case 39 bus test system for Scenario B.}\label{fig:const_case39_caseB}
	\vspace{-0.1cm}
\end{figure}

\begin{figure}[t]
	\hspace{-0.0cm}\subfloat{\includegraphics[keepaspectratio=true,scale=0.51]{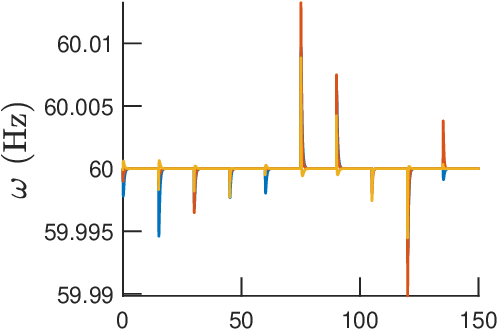}}{}{}\hspace{-0.15cm}
	\subfloat{\includegraphics[keepaspectratio=true,scale=0.51]{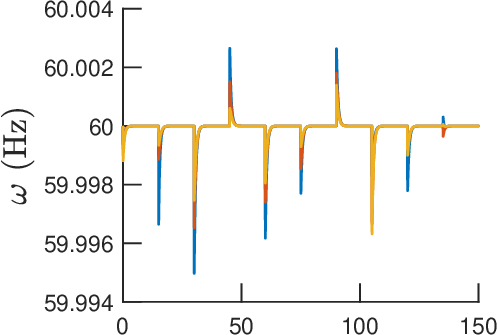}}{}{}\vspace{0.19cm}
	\hspace{-0.15cm}\subfloat{\includegraphics[keepaspectratio=true,scale=0.51]{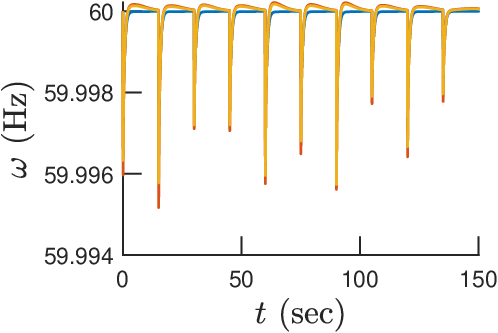}}{}{}\hspace{-0.16cm}
	\subfloat{\includegraphics[keepaspectratio=true,scale=0.51]{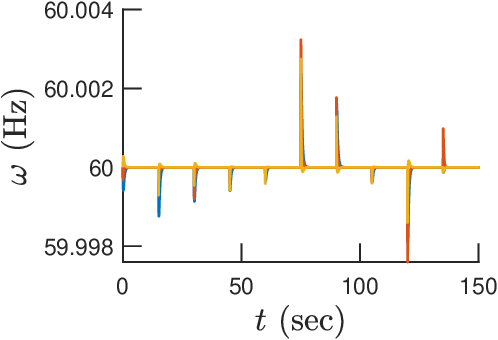}}{}{}\vspace{0.19cm}
	\caption{Generator frequencies under ten random disturbances in load and renewables for case 9, case 14, case 39, and case 57 test systems respectively.}\label{fig:omega_caseB}
	\vspace{-0.1cm}
\end{figure}

\begin{figure}[t]
	\hspace{-0.0cm}\subfloat{\includegraphics[keepaspectratio=true,scale=0.51]{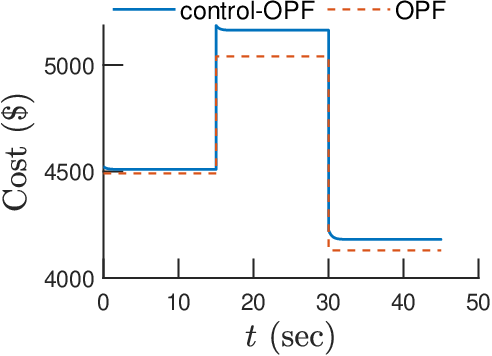}}{}{}\hspace{-0.1cm}
	\subfloat{\includegraphics[keepaspectratio=true,scale=0.51]{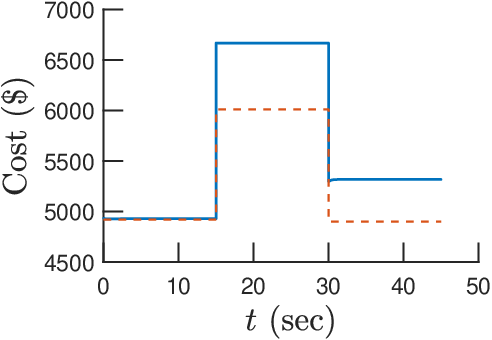}}{}{}\vspace{-0.1cm}
	\subfloat{\includegraphics[keepaspectratio=true,scale=0.51]{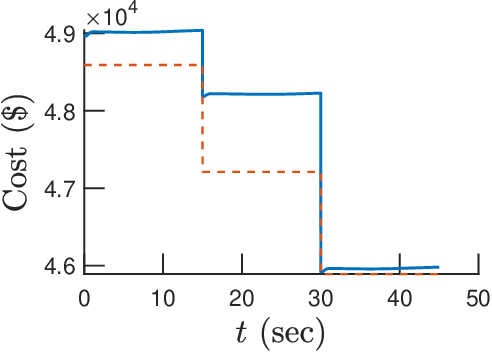}}{}{}\hspace{-0.0cm}
	\subfloat{\includegraphics[keepaspectratio=true,scale=0.51]{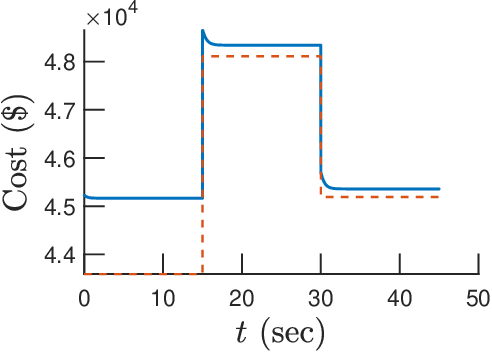}}{}{}\vspace{-0.19cm}
	\caption{Comparison of the operating cost of the system with control-OPF and OPF under Scenario A, case 9, case 14, case 39, and case 57, respectively.}\label{fig:cost}
	\vspace{-0.1cm}
\end{figure}

\begin{figure}[t]
	\hspace{-0.0cm}\subfloat{\includegraphics[keepaspectratio=true,scale=0.51]{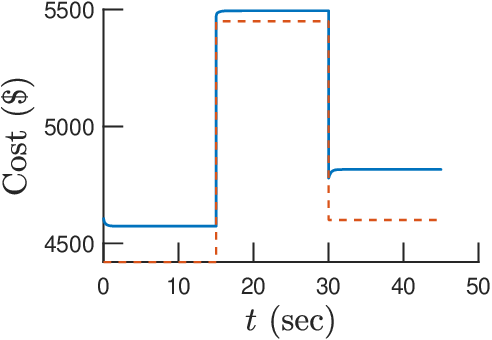}}{}{}\hspace{-0.1cm}
	\subfloat{\includegraphics[keepaspectratio=true,scale=0.51]{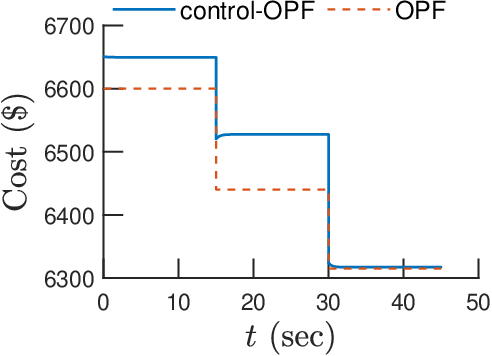}}{}{}\vspace{-0.1cm}
	\subfloat{\includegraphics[keepaspectratio=true,scale=0.51]{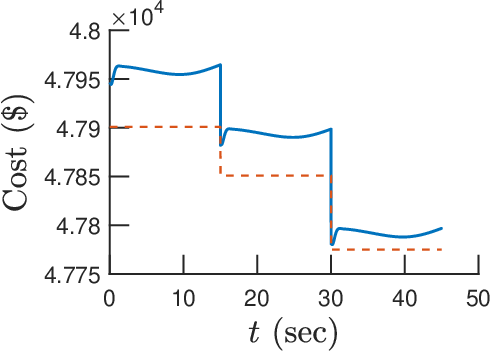}}{}{}\hspace{-0.1cm}
	\subfloat{\includegraphics[keepaspectratio=true,scale=0.51]{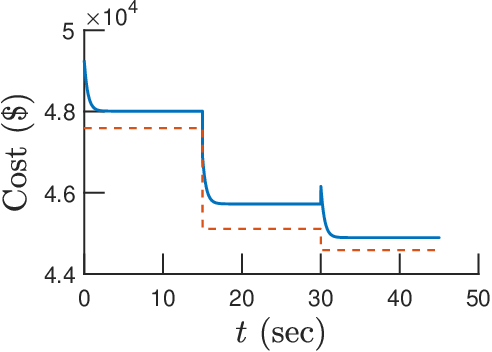}}{}{}\vspace{0.19cm}
	\caption{Comparison of the operating cost of the system with control-OPF and OPF under Scenario B, case 9, case 14, case 39, and case 57, respectively.}\label{fig:costB}
\end{figure}

\begin{figure}[t]
	\hspace{-0.0cm}\subfloat{\includegraphics[keepaspectratio=true,scale=0.5]{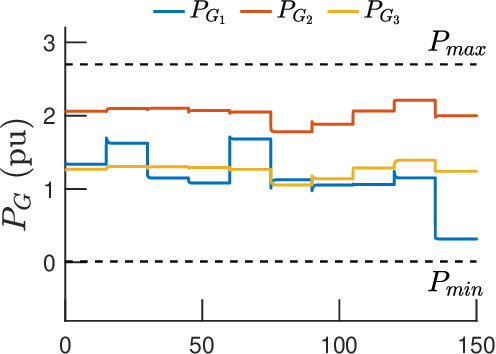}}{}{}\hspace{-0.15cm}
	\subfloat{\includegraphics[keepaspectratio=true,scale=0.5]{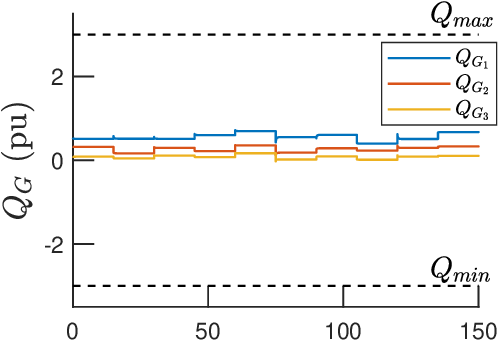}}{}{}\vspace{0.19cm}
	
	\subfloat{\includegraphics[keepaspectratio=true,scale=0.51]{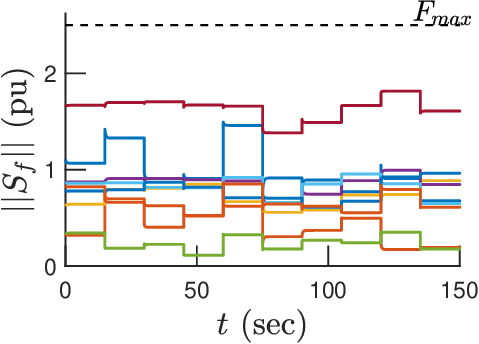}}{}{}\hspace{-0.1cm}
	\subfloat{\includegraphics[keepaspectratio=true,scale=0.49]{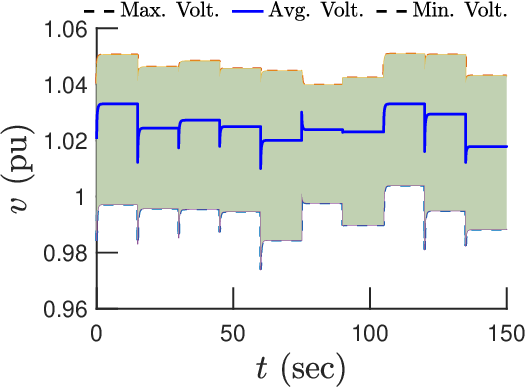}}{}{}\vspace{0.19cm}
	
	\caption{Active and reactive power generated by all the generators and their respective limits, line flows and their maximum rating, and the overall modulus of all bus voltages for case 9 bus test system for Scenario B.}\label{fig:const_case9_caseB}
\end{figure}

\begin{table*}[t]
	\centering
	\caption{Cost comparison for the control-OPF and ACOPF for Scenario A. The Control-OPF provides realtime frequency regulations and thus the cost is slightly higher as compared to ACOPF. All the costs shown in the table are in thousands of dollars ($\times$$10^3$ \$).}
	\label{tab:cost_caseA}
	\begin{tabular}{|c|c|c|c|c|c|}
		\hline
		Test System &
		Method &
		\begin{tabular}[c]{@{}c@{}} Total System  Cost \end{tabular} &
		\begin{tabular}[c]{@{}c@{}} Realtime System \\ Regulation Cost\end{tabular} & \begin{tabular}[c]{@{}c@{}} Maximum Deviation of \\ Frequency Nadir (pu) \end{tabular} & \begin{tabular}[c]{@{}c@{}} Percentage Improvement\\ as Compared to ACOPF \end{tabular} \\ \hline
		\multirow{2}{*}{Case 9}  & ACOPF  & 5.4188 & --- & 2.30 $\times$$10^{-3}$  & --- \\
		&  control-OPF & 5.5801 & 0.1613  & 2.01 $\times$$10^{-3}$  & 8.660 \\ \hline
		\multirow{2}{*}{Case 14} & ACOPF  & 8.4591 & --- & 1.72 $\times$$10^{-3}$   & --- \\
		&  control-OPF & 9.0121 & 0.5531 &  1.38 $\times$$10^{-3}$   & 19.77 \\ \hline
		\multirow{2}{*}{Case 39} & ACOPF  & 41.819 & --- &  -2.411 $\times$$10^{-3}$   & ---\\
		& control-OPF & 46.105 & 4.286 & -2.001 $\times$$10^{-3}$  & 17.01 \\ \hline
		\multirow{2}{*}{Case 57} & ACOPF  & 42.771 & --- &  0.109 $\times$$10^{-3}$   & ---\\
		&  control-OPF & 47.511 & 4.736 &  0.078 $\times$$10^{-3}$  & 28.44\\ \hline
	\end{tabular}
\end{table*}

\begin{table*}[]
	\centering
	\caption{Cost comparison for the control-OPF and ACOPF for Scenario B. Similar to Scenario A the control-OPF improves system transient stability by adding damping and thus the cost is slightly higher. All the costs shown in the table are in thousands of dollars ($\times$$10^3$ \$)}
	\label{tab:cost_cost_caseB}
	\begin{tabular}{|c|c|c|c|c|c|}
		\hline
		Test System &
		Method &
		\begin{tabular}[c]{@{}c@{}}Total System  Cost \end{tabular} &
		\begin{tabular}[c]{@{}c@{}} Realtime System \\ Regulation Cost\end{tabular} & \begin{tabular}[c]{@{}c@{}} Maximum Deviation of \\ Frequency Nadir (pu) \end{tabular} & \begin{tabular}[c]{@{}c@{}} Percentage Improvement \\ as Compared to ACOPF \end{tabular} \\ \hline
		\multirow{2}{*}{Case 9}  & ACOPF  & 6.319 & --- & 1.112 $\times$$10^{-3}$  & --- \\
		&  control-OPF & 6.410 & 0.091  & 0.982 $\times$$10^{-3}$  & 11.691 \\ \hline
		\multirow{2}{*}{Case 14} & ACOPF  & 10.431 & --- & -1.107 $\times$$10^{-3}$   & --- \\
		&  control-OPF & 12.522 & 2.091 &  -0.891 $\times$$10^{-3}$   & 19.513 \\ \hline
		\multirow{2}{*}{Case 39} & ACOPF  & 50.929 & --- &  1.136 $\times$$10^{-3}$   & ---\\
		& control-OPF & 60.001 & 5.072 & 0.803 $\times$$10^{-3}$  & 29.313 \\ \hline
		\multirow{2}{*}{Case 57} & ACOPF  & 44.771 & --- &  -0.181 $\times$$10^{-3}$   & ---\\
		&  control-OPF & 48.001 & 3.230 &  -0.142 $\times$$10^{-3}$  & 21.546\\ \hline
	\end{tabular}
\end{table*}

\subsection{Scenario A: Uncertainty in Load Demand}
In this section, we analyze the overall system cost with the control-OPF and compare it with the cost obtained by running  ACOPF under random disturbances in load demand. To that end, the simulations are carried out as follows: Initially, the system operates under steady-state conditions, meaning the overall demand is exactly equal to the power generated by load and renewables. Thus there are no transients in the system and the system rests in an equilibrium state. Then right after $t>0$ ten random (with varying uncertainty) step disturbances in load demand have been added as follows: $P'_d + Q'_d =(1+\delta_d)(P^0_d + Q^0_d) + w_d(t)$, where $\delta_d$ represent the amount of the disturbance, $ w_d(t)$ is a Gaussian noise with zero mean and variance of $0.01(P^0_d + Q^0_d)$, $P^0_d$, $Q^0_d$ are the initial active and reactive load demand, and $P'_d$, $Q'_d$ is the new load demand after the disturbances has been applied. For every ten simulations, the value for $\delta_d$ is selected randomly in $[0.01,0.08]$ for case 9 and case 14, for case 39 the range is chosen in $[0.001,0.02]$, while for case 57 $\delta_d$ is randomly picked in $[0.001,0.01]$.

After the disturbance, the power system is stabilized via control-OPF, and the gain $\m K$ which is computed offline. Notice that for every load disturbance, we get time-varying generator power output vector $\mP_G$ and $\mQ_G$. The vector $\mP_G$ is then plugged into the quadratic cost equation of the generators (given in MATPOWER) and finally average is taken to compute the final cost. In this way, we get the overall cost of the system with the control-OPF acting in realtime to redistribute the power from the generator in response to the disturbances. For similar uncertainties in load demand, OPF is also carried out ten times, and an average of the overall cost is computed to determine the system cost for random loads with OPF.

To that end, a comparison of the overall system cost with control-OPF and OPF for this case study is presented in Tab. \ref{tab:cost_caseA}. We can note that for different test networks, the average cost of system operation under various load disturbances is close to the average cost computed via just running OPF. This can also be corroborated from Fig. \ref{fig:cost} from which we can see that the cost of control-OPF is close to the cost obtained from OPF for case 9 and case 14 test systems. The extra cost incurred in the case of control-OPF can be seen as the system regulation cost. As seen from Tab. \ref{tab:cost_caseA} with the control-OPF there is around $10-30\%$ improvement in the frequency nadir for various test systems resulting in improved system transient stability. This is because the proposed approach makes sure that the power system quickly converges to its equilibrium conditions and is $\mathcal{H}_\infty$-stable after a large disturbance.

Fig. \ref{fig:Pg Qg all caseA} also illustrate the time-varying power generation set-points (for the first three simulations) generated by control-OPF and static set-points received via solving OPF and we can observe that both of them are not far away from each other.  Moreover, in Fig. \ref{fig:const_case9_caseA} we present active and reactive power from all the generators, line flows, and modulus of bus voltages for the case 9 test system. Notice that, line flows are computed from the state vectors as follows:
\begin{align*}
	\m S_f = [\mC_f \mV]\mY^*_f\mV^*,  \,\,\,\,\,	\m S_t = [\mC_t \mV]\mY^*_t\mV^*
\end{align*}
where $\m S_f$,  $\m S_t$ are apparent power flows from both ends (from bus and to bus) of the transmission line respectively, $\mV$ are the bus voltages,  $\mY^*_f$, $\mY^*_t$ represent the conjugate of \textit{from} and \textit{to} bus admittance matrices, while  $\mC_f$, $\mC_t$ are binary matrices and it generates all \textit{from} and \textit{to} end buses of the transmission lines.

With that in mind, we can clearly see from Fig. \ref{fig:const_case9_caseA} that all the line flows, bus voltages, and generator's power outputs are within their prescribed limits and thus the control-OPF successfully satisfies all the system constraints that are usually modeled in OPF. Similarly for all the other test systems, we can see from Tab. \ref{tab:contratints_caseA} that the maximum instantaneous value for the line flows, and active and reactive power generations are less than their respective maximum limits. Thus the proposed control-OPF satisfies the constraints of the system---and no ACOPF constraint violations are incurred.


\begin{table*}[]
	\vspace{-0.5cm}
	\centering
	\caption{Summary of ACOPF constraints for different test systems with control-OPF for Scenario A. The results indicate \textbf{no} constraint violations for flows, maximum active/reactive powers.}
	\label{tab:contratints_caseA}
	\begin{tabular}{|c|c|c|c|c|c|}
		\hline
		System & \begin{tabular}[c]{@{}c@{}}$\max_t$(${\mS_f(t)}\hspace{-0.05cm} -\hspace{-0.05cm} \mS_{\mr{max}}$)\\ (pu)\end{tabular} & \begin{tabular}[c]{@{}c@{}}$\max_t$(${\mS_t(t)}\hspace{-0.05cm}-\hspace{-0.05cm}\mS_\mr{max})$\\ (pu)\end{tabular} & \begin{tabular}[c]{@{}c@{}}$\max_t$(${\mP_g(t)}\hspace{-0.05cm} -\hspace{-0.05cm} {\mP_\mr{max}}$)\\ (pu)\end{tabular} & \begin{tabular}[c]{@{}c@{}}$\max_t$(${\mQ_g(t)}\hspace{-0.05cm}-\hspace{-0.05cm}{\mQ_\mr{min}}$)\\ (pu)\end{tabular} & \begin{tabular}[c]{@{}c@{}}$\max_t$(${\mQ_g(t)}\hspace{-0.05cm}-\hspace{-0.05cm}{\mQ_\mr{max}}$)\\ (pu)\end{tabular} \\ \hline
		Case 9      & -0.5120                                                                & -0.4401                                                             & -0.9631                                                                  & 3.1402                                                                & -0.1091                                                              \\ \hline
		Case 14     & -0.4101                                                                    & -0.2170                                                                 & -0.6706                                                                  & 0.1926                                                                & -0.1006                                                               \\ \hline
		Case 39     & -0.1763                                                                & -0.1695                                                             & -0.0918                                                                  & 2.1021                                                               & -0.1921                                                               \\ \hline
		Case 57     & -0.1391                                                                & -0.4112                                                             & -0.0111                                                                  & 2.1120                                                                & -1.0326                                                               \\ \hline
	\end{tabular}
\end{table*}

\begin{table*}[t]
	\centering
	\caption{Summary of ACOPF constraints for different test systems with the control-OPF for Scenario B. The results indicate \textbf{no} constraint violations for flows, maximum active/reactive powers.}\label{tab:Constraints_caseB}
	
	\begin{tabular}{|c|c|c|c|c|c|}
		\hline
		System & \begin{tabular}[c]{@{}c@{}}$\max_t$(${\mS_f(t)} \hspace{-0.05cm}-\hspace{-0.05cm} \mS_{\mr{max}}$)\\ (pu)\end{tabular} & \begin{tabular}[c]{@{}c@{}}$\max_t$(${\mS_t(t)}\hspace{-0.05cm}-\hspace{-0.05cm}\mS_{\mr{max}})$\\ (pu)\end{tabular} & \begin{tabular}[c]{@{}c@{}}$\max_t$(${\mP_g(t)} \hspace{-0.05cm}-\hspace{-0.05cm} {\mP_{\mr{max}}}$)\\ (pu)\end{tabular} & \begin{tabular}[c]{@{}c@{}}$\max_t$(${\mQ_g(t)}\hspace{-0.05cm}-\hspace{-0.05cm}{\mQ_{\mr{min}}}$)\\ (pu)\end{tabular} & \begin{tabular}[c]{@{}c@{}}$\max_t$(${\mQ_g(t)}\hspace{-0.05cm}-\hspace{-0.05cm}{\mQ_{\mr{max}}}$)\\ (pu)\end{tabular} \\ \hline
		Case 9      & -0.3009                                                                & -0.4110                                                             & -0.0401                                                                  & 1.2130                                                                & -1.990                                                               \\ \hline
		Case 14     & -0.1099                                                                    & -0.2101                                                                & -0.1099                                                                  & 1.2216                                                                & -0.0115                                                               \\ \hline
		Case 39     & -0.1421                                                                & -0.7321                                                             & -0.0091                                                                  & 0.2510                                                                & -0.0109                                                              \\ \hline
		Case 57     & -0.7020                                                                & -0.1981                                                             & -0.0021                                                                  & 2.01910                                                               & -0.0307                                                               \\ \hline
	\end{tabular}
	\vspace{-0.3cm}
\end{table*}
\subsection{Scenario B: Uncertainty in Renewable Power Generation}
Here we analyze the cost of operating the system with control-OPF and compare it with OPF under random uncertainty in renewable power generations. To that end, the simulations in this section are performed as follows: Initially, the power generation from renewables is $\mP^0_R$, $\mQ^0_R$, then right after $t>0$, a random disturbance has been added and the power output from renewables are given as: $P'_R + Q'_R =(1+\delta_R)(P^0_R + Q^0_R) + w_R(t)$, where $\delta_R$ represent the severity of the disturbance, $\m w_R(t)$ is the random noise as shown in Fig. \ref{fig:random_renew}, and $P'_R$, $Q'_R$ are the updated power output from renewables after the disturbance. With that in mind, we carry out ten simulations and for each simulation, the value for $\delta_R$ is selected randomly in $[-0.01,0.03]$ for case 9 and case 14, for case 39 it is in $[-0.001,0.01]$, while for case 57 it is chosen randomly in $[-0.01,0.02]$. 

To that end, from Tab. \ref{tab:cost_cost_caseB} we can see that the difference between system operating cost with control-OPF and by just running OPF are close to each other. Again the extra cost observed in the case of control-OPF is the system regulation cost and results in the improvement of system transient stability for all the test systems. This means that control-OPF not only ensures transient stability of the system via realtime feedback---which can also be verified from Fig. \ref{fig:omega_caseB}, we can see that all the generator frequencies quickly converge to their equilibrium after large disturbance, but also makes sure that the power redistribution from the generators after a disturbance is such that it is close to OPF cost. In Fig. \ref{fig:const_case39_caseB} we also illustrate for the 39-bus test system the active power output of generator 2, the reactive power output of generators 1 to 4, line flows of a couple of transmission lines, and modulus of bus voltages for all ten simulations. We can clearly see that for every random renewable uncertainty, the generator power output, line flows, and bus voltages are within their prescribed limits. Thus ensuring that the system constraints are satisfied.

These results can also be corroborated from Tab. \ref{tab:Constraints_caseB}, from which we can observe that for all test systems, the instantaneous active/reactive power outputs and transmission \textit{to} and \textit{from} line flows are less/greater than their respective maximum/minimum limits. Notice that the reason it satisfies all the system constraints is because the proposed control-OPF makes sure that the system is stable (in terms of $\mathcal{H}_\infty$ and Lyapunov stability) and it inherently encodes the algebraic constraints (power balance and generator stator constraints) of power system in its feedback control architecture. 

\begin{figure}[]
	\hspace{-0.0cm}\subfloat{\includegraphics[keepaspectratio=true,scale=0.5]{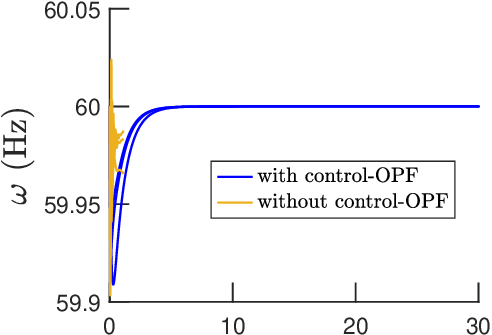}}{}{}\hspace{-0.15cm}
	\subfloat{\includegraphics[keepaspectratio=true,scale=0.5]{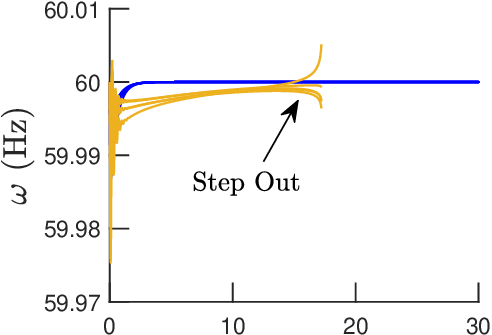}}{}{}\vspace{0.19cm}
	
	\subfloat{\includegraphics[keepaspectratio=true,scale=0.51]{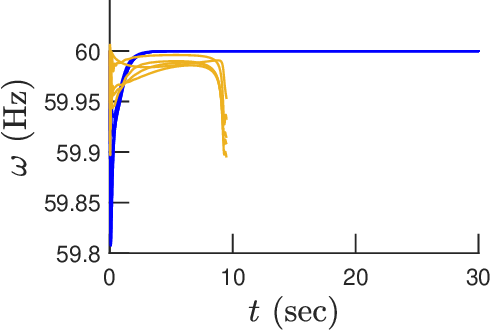}}{}{}\hspace{-0.1cm}
	\subfloat{\includegraphics[keepaspectratio=true,scale=0.49]{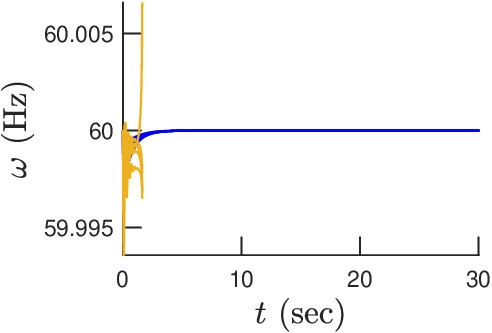}}{}{}\vspace{0.19cm}
	
	\caption{The generator frequencies for 9-bus (top-left), 14-bus (top-right), 39-bus
		(bottom-left), and 57-bus (bottom-right) test systems, for disturbance in load demand
		and renewable power.}\label{fig:high_disturbance}
\end{figure}

Furthermore, it is worth mentioning here that besides improving frequency nadir, the control-OPF also makes the power system more robust toward various uncertainties from load and renewables. To verify this, we further increased the severity of load and renewable uncertainty by increasing the value of $\delta_d$ and $\delta_R$, respectively, and we simulated the systems without and with control-OPF under these disturbances. The results are presented in Fig. \ref{fig:high_disturbance}. We can see that for all the test systems, without control-OPF the system becomes unstable and loses its synchrony while with the proposed control-OPF the system remains stable and synchronized.

To that end, since the proposed control-OPF, ensures guaranteed stability, satisfies all the system constraints, and the overall system cost after a large disturbance in load and renewable is close to the cost obtained from OPF. Then the need to solve OPF after $5-10$ minutes in the tertiary layer (or economic dispatch layer) of the power system can be eliminated. This is because the control-OPF acts in realtime through feedback provided by the PMUs and it also ensures system stability as discussed in Sec. \ref{sec:NDAE_control}.

\section{Paper Summary, Limitations, and Future Work}\label{sec:conclusion}
In this work, we propose a new method to approximate the OPF problem using feedback control theory. The proposed algorithm namely control-OPF is based on Lyapunov stability and it explicitly models the algebraic constraints of the power system in the controller architecture. These algebraic constraints (especially the power balance equations) are part of the OPF problem, since the control-OPF inherently satisfies these constraints then the need for solving OPF after $5-10$ minutes in the tertiary layer of the power system can be rethought or potentially eliminated. 


Given the case studies, we present preliminary answers to the posed research questions \textit{Q1--Q4} posed in Section~\ref{sec:casestudies}. 
\begin{itemize}
	\item \textit{A1.} We observe that control-OPF approach yields a cost function that is on average slightly higher than the ACOPF under transient conditions. The extra cost incurred can be seen as system regulation cost. 
	\item \textit{A2.} The control-OPF produces more than just time-varying, realtime generator setpoints and deviations; it produces realtime regulation of the grid's voltages and frequencies. The slight extra cost of control-OPF results in on average $10-30\%$ improvement in frequency nadir, depending on the studied system and the assumed conditions. 
	\item \textit{A3.} The control-OPF approach results in no constraint violations for all studied power systems under different realizations of renewables, loads, and initial conditions. 
	\item \textit{A4.} While the OPF knows exact values for all uncertain loads and renewables (needed to compute OPF setpoints), the control-OPF is truly uncertainty-unaware. The former needs vectors of uncertainty from renewables and loads; the latter hedges against it.  Hence one could argue that the cost comparison is objectively unfair to the control-OPF. A fairer comparison would be with a stochastic OPF, which is also uncertainty-aware.
\end{itemize}
The limitations of the presented work are as follows: 
\begin{itemize}
	\item As compared to ACOPF the proposed method herein does not take into account the inequality constraints and the quadratic cost equations of synchronous generators.  
	\item The control-OPF formulation does not include theoretical near-optimality guarantees. It is only under computationally feasible conditions, such as the existence of an NDAE feedback controller gain matrix $\mK$, that the near-optimality performance of the proposed method can be empirically evaluated.
	\item The near-optimality performance of the control-OPF method can potentially deteriorate and can also result in some operational constraint violations under potentially different operating conditions.
\end{itemize}
Future work will focus on addressing the above limitations, comparing this framework with a robust version of ACOPF, extending the dynamic model to incorporate power-electronics-based models of renewable energy resources such as wind and solar farms, and investigating the performance of robust $\mathcal{H}_{2}$- or $\mathcal{L}_{\infty}$-based controllers in terms of costs and response to uncertainty.


\vspace{-0.3cm}
\bibliographystyle{IEEEtran}
\bibliography{bib_file}
\vspace{-0.3cm}

\end{document}